\documentclass[%
 reprint,
superscriptaddress,
 amsmath,amssymb,
 aps,
 prd,
floatfix,
]{revtex4-2}

\usepackage{graphicx}
\usepackage{dcolumn}
\usepackage{bm}
\usepackage{subfigure}

\usepackage{hyperref}

\usepackage{xspace}
\def\etal{\textit{et~al.}\xspace}

\begin{document}


\title{Energy spectrum and mass composition of the primary cosmic rays \\ based on the intensity of muon bundles detected in the NEVOD-DECOR experiment}

\author{A.G. Bogdanov}
 \email{agbogdanov@mephi.ru}
 \affiliation{National Research Nuclear University MEPhI (Moscow Engineering Physics Institute), Russia}
\author{A. Chiavassa}
 \affiliation{Dipartimento di Fisica Dell’ Università Degli Studi di Torino, Italy}
 \affiliation{Sezione di Torino Dell’ Istituto Nazionale di Fisica Nucleare – INFN, Italy}
\author{D.M. Gromushkin}
 \affiliation{National Research Nuclear University MEPhI (Moscow Engineering Physics Institute), Russia}
\author{S.S. Khokhlov}
 \affiliation{National Research Nuclear University MEPhI (Moscow Engineering Physics Institute), Russia}
\author{V.V. Kindin}
 \affiliation{National Research Nuclear University MEPhI (Moscow Engineering Physics Institute), Russia}
\author{\\K.G. Kompaniets}
 \affiliation{National Research Nuclear University MEPhI (Moscow Engineering Physics Institute), Russia}
\author{A.Yu. Konovalova}
 \affiliation{National Research Nuclear University MEPhI (Moscow Engineering Physics Institute), Russia}
\author{K.I. Mannanova}
 \affiliation{National Research Nuclear University MEPhI (Moscow Engineering Physics Institute), Russia}
\author{G. Mannocchi}
 \affiliation{Osservatorio Astrofisico di Torino – INAF, Italy}
\author{A.A. Petrukhin}
 \affiliation{National Research Nuclear University MEPhI (Moscow Engineering Physics Institute), Russia}
\author{G. Trinchero}
 \affiliation{Sezione di Torino Dell’ Istituto Nazionale di Fisica Nucleare – INFN, Italy}
 \affiliation{Osservatorio Astrofisico di Torino – INAF, Italy}
\author{\\I.Yu. Troshin}
 \affiliation{National Research Nuclear University MEPhI (Moscow Engineering Physics Institute), Russia}
\author{I.A. Shulzhenko}
 \affiliation{National Research Nuclear University MEPhI (Moscow Engineering Physics Institute), Russia}
\author{V.V. Shutenko}
 \affiliation{National Research Nuclear University MEPhI (Moscow Engineering Physics Institute), Russia}
\author{V.S. Vorobev}
 \affiliation{National Research Nuclear University MEPhI (Moscow Engineering Physics Institute), Russia}
\author{I.I. Yashin}
 \affiliation{National Research Nuclear University MEPhI (Moscow Engineering Physics Institute), Russia}
\author{E.A. Yurina}
 \affiliation{National Research Nuclear University MEPhI (Moscow Engineering Physics Institute), Russia}


\begin{abstract}
The results of the analysis of the NEVOD-DECOR data on the study of inclined muon bundles \\ ($\theta$ = 40 - 85$^{\circ}$) of cosmic rays for the period from 2012 to 2023 are presented. An original method for studying the muon component of extensive air showers, local muon density spectra, was used. The data are compared with the calculations based on the simulation of air showers using the CORSIKA program for different models of hadronic interactions. The estimates of the energy spectrum and the behavior of the mass composition of primary cosmic rays in a wide energy range from $2 \times 10^{15}$ to $3 \times 10^{18}$ eV were obtained. They are compared with the data of other experiments.
\end{abstract}

\maketitle


\section{Introduction}

Primary cosmic rays (CRs) carry important information about physical processes occurring in our Galaxy and in the Universe. Verification of hypotheses about the origin, acceleration and propagation of primary CRs is directly related to the study their energy spectrum, mass composition and arrival directions (anisotropy). To date, the only way to study the properties of the primary CR flux in the energy range above $10^{15}$ eV is to detect extensive air showers (EAS) on the Earth's surface. An extensive air shower is a nuclear-electromagnetic cascade that consists of many secondary particles generated in the interaction of primary CRs with the nuclei of air, as well as in result of subsequent interactions and decays of particles in the Earth's atmosphere. Thus, when interpreting the data on the detection of EAS along with the energy spectrum and mass composition of primary CRs, the characteristics of hadronic interactions play an important role.

The EAS muon component is formed mainly in the decays of charged $\pi$- and \textit{K}-mesons and is a traditional tool for studying the composition of CRs, as well as for validating models of hadronic interactions at the ultra-high energies. Currently, one of the most urgent problems in the Cosmic Ray Physics is the ``muon puzzle". In a number of experiments, an excess of muons with respect to the theoretical predictions even under the assumption of an extremely heavy (iron group nuclei) composition of CRs at energies of $\sim 10^{18}$ eV and higher is observed.

The events with muon bundles, which represent a simultaneous (within tens of nanoseconds) passage of several penetrating particles with almost parallel tracks through the detector, are of particular interest. The typical energies of muons in bundles are several times greater than those for air showers as a whole. Such muons originate from high-energy hadrons (probably genetically related to each other) generated in the forward kinematic region of particle interactions in the atmosphere.

In this paper, the results of our long-term experiment on the systematic investigation of muon bundles in a wide range of zenith angles are presented. The uniqueness of the NEVOD-DECOR setup, compared to numerous experiments in which the EAS muon component is being studied today, is due to the combination of a vertically oriented track detector and a Cherenkov water calorimeter. This makes it possible to study inclined muon bundles (up to the horizon), as well as to measure their energy deposit.

\section{NEVOD-DECOR setup}

The NEVOD-DECOR setup is located in a dedicated laboratory building at MEPhI (Moscow, Russia) and includes a Cherenkov water calorimeter NEVOD with a volume of $26 \times 9 \times 9$ m$^{3}$ and a track detector DECOR with an area of 70 m$^{2}$, which are combined by a common trigger system.

\begin{figure*}[ht]
 \begin{minipage}[h]{0.475\textwidth}
  \includegraphics[width=0.65\textwidth]{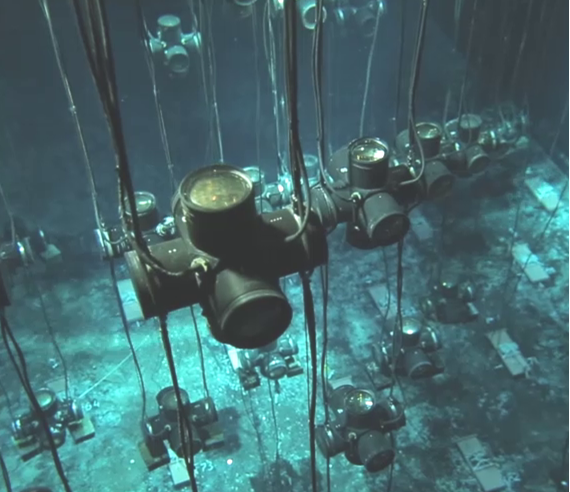}
 \end{minipage}
 \begin{minipage}[h]{0.475\textwidth}
  \includegraphics[width=0.55\textwidth]{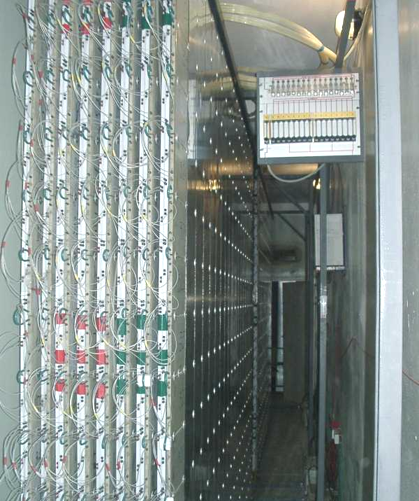}
 \end{minipage}
\caption{\label{fig:1} The quasi-spherical measuring modules of the detecting system of the Cherenkov water calorimeter NEVOD (left). The supermodules of the track detector DECOR mounted in one of the galleries of the laboratory building (right).}
\end{figure*}

The multipurpose Cherenkov water detector NEVOD~\cite{Petrukhin2015} (Fig.~\ref{fig:1}, left) consists of 91 quasi-spherical measuring modules (QSMs), which are located at the nodes of a spatial lattice with steps of 2.5 m along the detector, 2 m across the detector and in its depth. Each QSM consists of six photomultipliers (PMTs) with a flat photocathode oriented along the axes of an orthogonal coordinate system. This design ensures almost the same efficiency of detecting Cherenkov radiation from any direction. A wide dynamic range of each PMT (from 1 to $10^5$ photoelectrons) is achieved by the use of two-dynode signal readout. This allows calorimetric studies, in particular, measuring the energy deposit of muon bundles~\cite{Yurina2021}.

\vspace{10pt}
The track detector DECOR~\cite{Barbashina2000} (Fig.~\ref{fig:1}, right) was constructed within the framework of Russian-Italian collaboration and is intended for detection of multi-muon events on the Earth's surface in a wide range of zenith angles. It consists of 8 vertical supermodules (SMs), which are mounted in three building galleries around the Cherenkov water calorimeter NEVOD. Each SM is an assembly of 8 parallel planes of plastic gas discharge tubes (3.5 m long and $9 \times 9$ mm$^{2}$ in cross-section) operating in a limited streamer mode. The effective area of the SM is 8.4 m$^{2}$. The system of external strips on each SM plane provides reading out of induced signals in two coordinates (256 \textit{X}-channels along the tubes and 256 \textit{Y}-channels across them). The spatial and angular accuracy of reconstruction of muon tracks crossing the SM is better than 1 cm and 1$^{\circ}$, respectively.

\begin{figure}[h!]
 \begin{minipage}[h]{0.475\textwidth}
  \includegraphics[width=.9\linewidth]{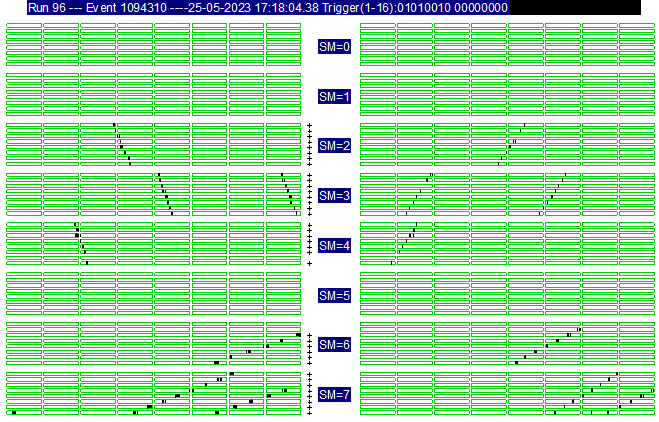}
 \end{minipage}
 \begin{minipage}[h]{0.475\textwidth}
  \includegraphics[width=.9\linewidth]{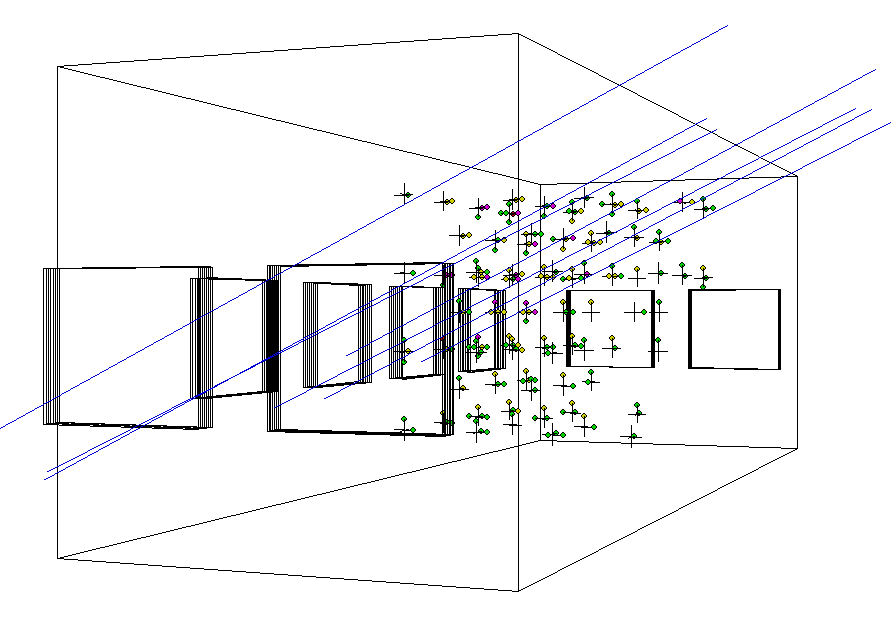}
 \end{minipage}
\caption{\label{fig:2} The example of a typical event with a muon bundle detected by the NEVOD-DECOR. The DECOR response (8 SMs in two projections) is shown in the top panel. The geometric reconstruction of the muon tracks in this event, as well as the scheme of the NEVOD-DECOR setup are shown in the bottom panel.}
\end{figure}

\begin{figure*}[t]
 \begin{minipage}[h]{0.475\textwidth}
  \includegraphics[width=.95\linewidth]{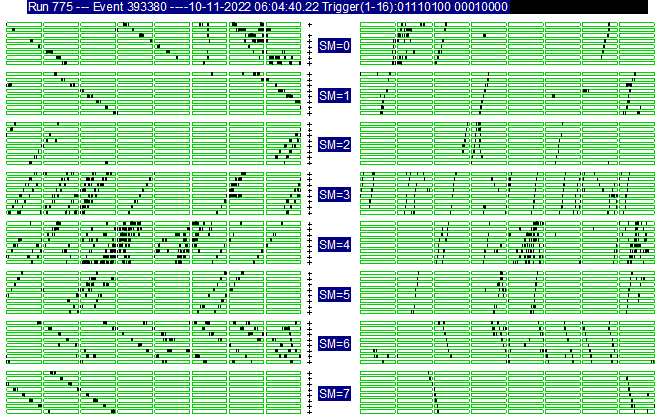}
 \end{minipage}
 \begin{minipage}[h]{0.475\textwidth}
  \includegraphics[width=.95\linewidth]{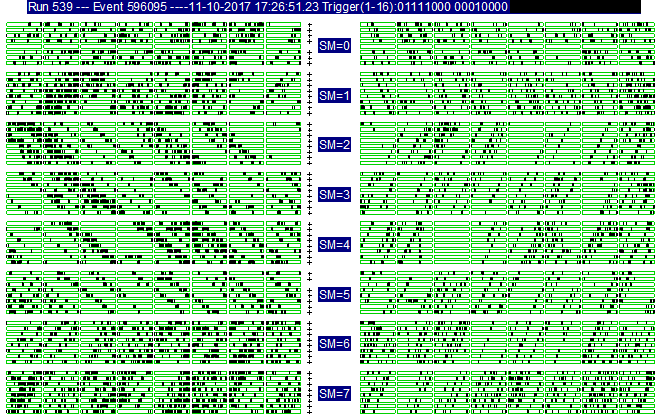}
 \end{minipage}
\caption{\label{fig:3} The examples of record-breaking multi-muon events detected by the NEVOD-DECOR setup (response of the detector DECOR is shown) with large zenith angles and large multiplicities.}
\end{figure*}

\section{Experimental data}

Previously, the results of the analysis of the first data from the NEVOD-DECOR experiment on the study of muon bundles in period 2002-2007, as well as in combined periods 2002-2007 and 2012-2016 data were published in~\cite{Bogdanov2010,Bogdanov2018}.

In this paper, we used the NEVOD-DECOR data accumulated over five series of data-taking carried out from May 2012 to July 2023 (the ``live'' observation time is 75238 hours). For this period, 129173 events with muon multiplicity $m \geq 5$ and zenith angles $\theta \geq 55^{\circ}$ were selected.

Additionally, to expand the primary particle energy range under study towards the lower energies, 30375 events (6324 hours) with muon multiplicity $m \geq 5$ and 4130 events (1043 hours) with multiplicity $m = 4$ were selected from the part of experimental data corresponding to the smaller zenith angles $\theta =$ 40 – 55$^{\circ}$.

An example of a typical event with a muon bundle ($m = 7$ and $\theta \approx 62^{\circ}$) detected by the NEVOD-DECOR setup is shown in Fig.~\ref{fig:2}. The top panel of the figure shows the response of eight supermodules of the DECOR in two projections (\textit{Y} and \textit{X}). A spatial reconstruction of the same event, as well as a diagram of the setup are shown in the bottom panel of the figure. Thin lines show muon tracks reconstructed according to the DECOR data, small circles show the triggered PMTs in the NEVOD calorimeter (colors reflect signal amplitudes), and large rectangles show the DECOR supermodules. It should be noted that multi-muon events most often have a bright signature in the track detector, and their interpretation is almost unambiguous.

The capabilities of the NEVOD-DECOR experiment to study CRs at ultra-high energies are limited mainly by the statistics at large zenith angles (10 events with $\theta \geq 85^{\circ}$ during the observation time) and the spatial resolution at high multiplicities ($\approx 3$ cm for two tracks). The examples of record-breaking muon bundles that were excluded from the data analysis for these reasons are shown in Fig.~\ref{fig:3}. The left panel shows an event with $m = 33$ and $\theta \approx 84^{\circ}$ (the estimate of primary particle energy is $E_\text{0} \approx 4\times10^{19}$ eV). The right one shows an event with $m = 124$ and $\theta \approx 75^{\circ}$ ($E_\text{0} \approx 7\times10^{18}$ eV). The lower limit on the energy of primary particles is due to the low density of muons in the detected air showers. 

\begin{figure}[h]
 \begin{minipage}[h]{0.475\textwidth}
  \includegraphics[width=.775\linewidth]{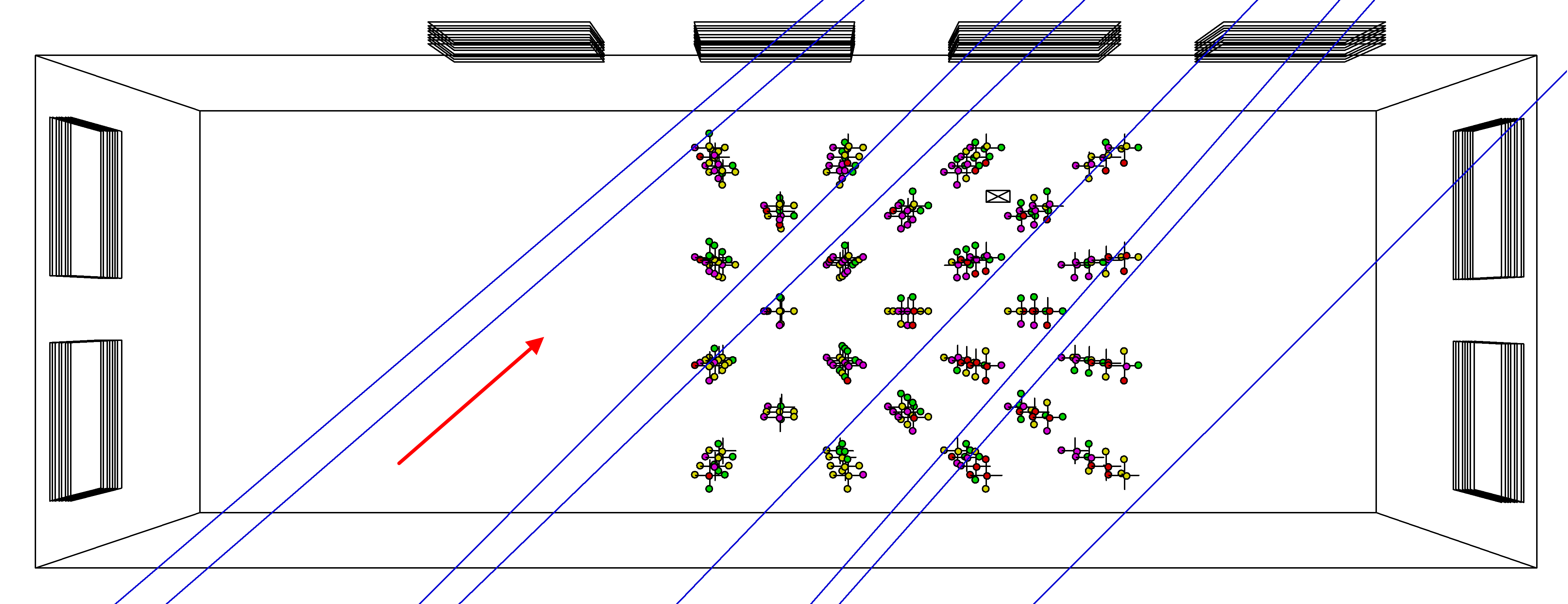}
 \end{minipage}
 \begin{minipage}[h]{0.475\textwidth}
  \includegraphics[width=.775\linewidth]{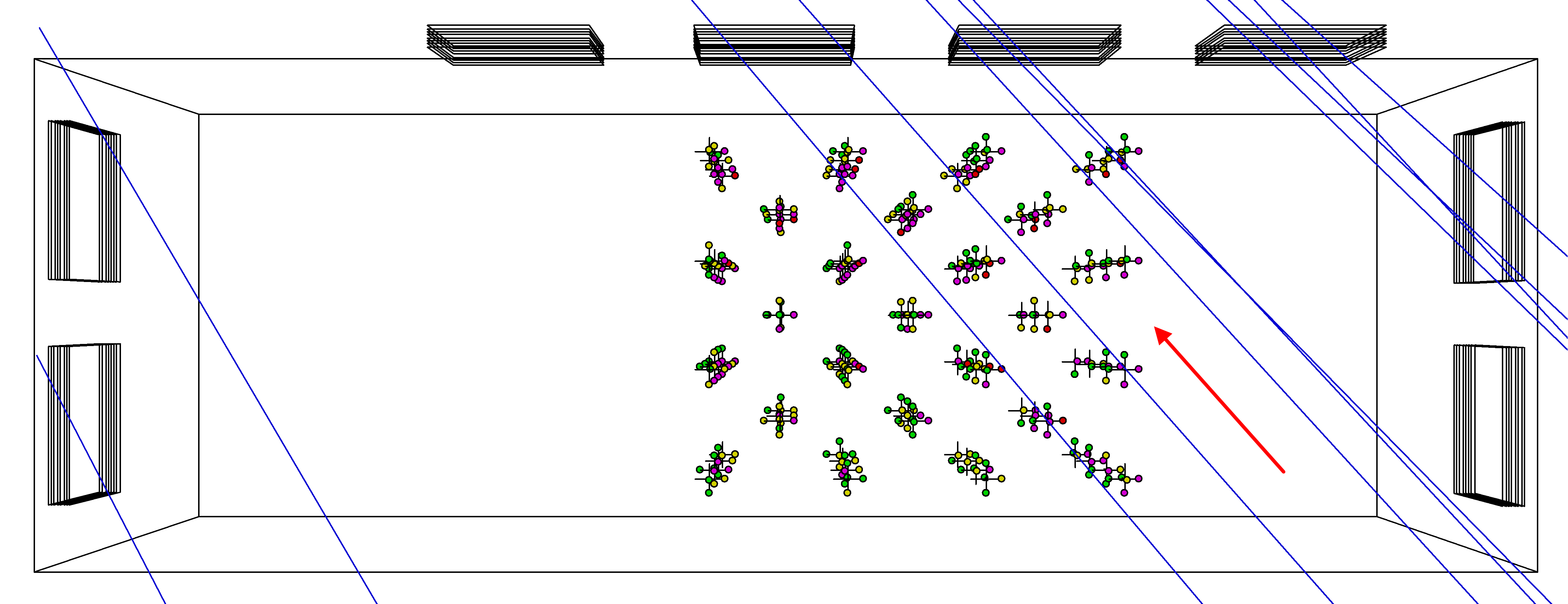}
 \end{minipage}
\caption{\label{fig:4} The examples of geometric reconstruction of events with muon bundles selected in two sectors of azimuthal angles with a width of $60^{\circ}$ each. Figure shows the top view of the NEVOD-DECOR scheme.}
\end{figure}

At moderate zenith angles, the identification of muon tracks can be complicated due to the residual contribution of the electron-photon and hadron components of the air showers, which have not yet been completely absorbed in the atmosphere. To eliminate their influence, the muon bundles were selected in two $60^{\circ}$-wide sectors of azimuthal angles where six of the eight DECOR SMs are shielded by the water volume of the NEVOD calorimeter (Fig.~\ref{fig:4}). Then, data from only these six SMs were used to estimate the number of tracks in the muon bundle. To ensure the same selection conditions, this constraint was used throughout the studied range of zenith angles. In this case, the threshold energy of muons is about 2 GeV.

The procedure for selecting events with muon bundles hitting the DECOR track detector was carried out in several stages:

- hardware selection of 3-fold coincidences of signals from different SMs within a time gate of 250 ns, hits of any two even and two odd planes of each SM are required, the frequency of such events is about 0.25 s$^{-1}$;

- software selection and reconstruction of candidate events containing quasi-parallel tracks within a $5^{\circ}$ cone (i.e., straight lines passing through at least five fired channels in different SM planes);

- final classification of events and visual track counting by several operators.

\begin{figure}[ht]
\includegraphics[trim={1.5cm 1.5cm 1.5cm 1.5cm}, clip, width=.9\linewidth]{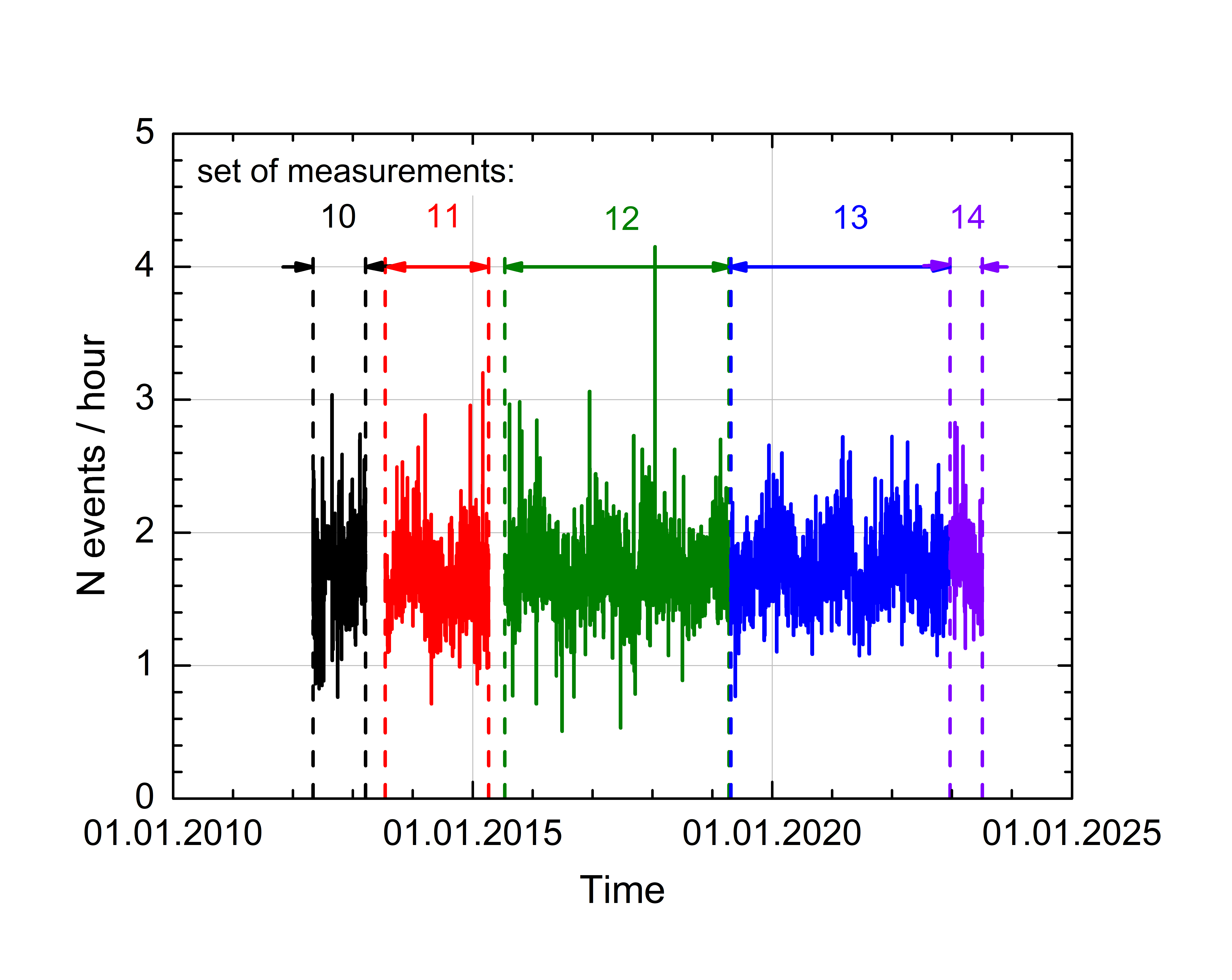}
\caption{\label{fig:5} The time dependence of the muon bundle counting rate of the NEVOD-DECOR (different series of data-taking are shown).}
\end{figure}

Express data analysis, which was carried out synchronously with data acquisition during the entire experiment, ensured quick identification and troubleshooting of equipment. Fig.~\ref{fig:5} shows the dependence of the muon bundle detection frequency ($m \geq 5$, $\theta \geq 55^{\circ}$) on time, which confirms stable operation of the NEVOD-DECOR setup for many years. The average counting rate was about 1.7 hour$^{-1}$, and its variations were due to barometric and temperature effects~\cite{Yurina2019}.

\begin{figure}[h!]
 \begin{minipage}[h]{0.475\textwidth}
  \includegraphics[trim={1.5cm 1cm 1.7cm 1.5cm}, clip, width=.9\linewidth]{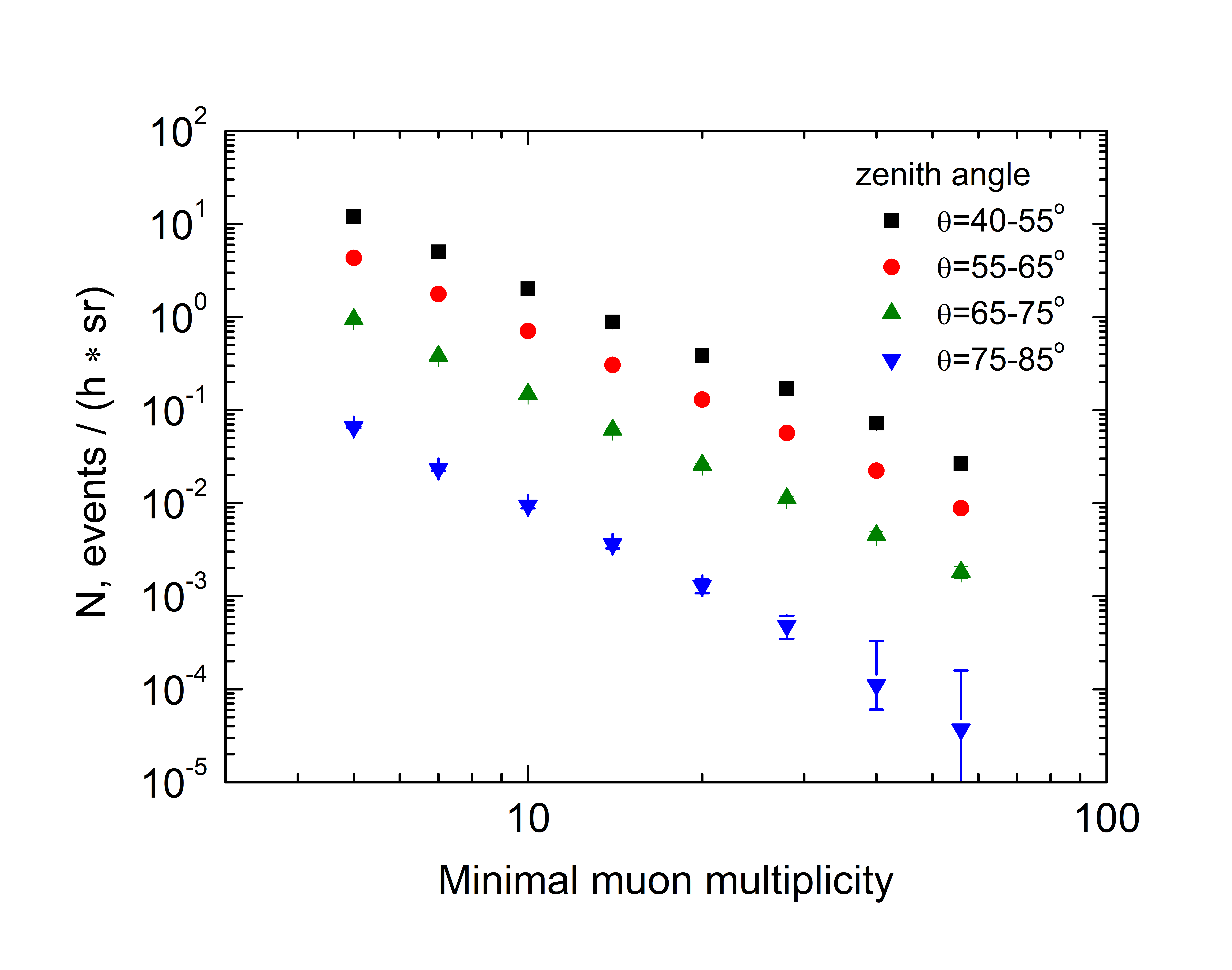}
 \end{minipage}
 \begin{minipage}[h]{0.475\textwidth}
  \includegraphics[trim={1.5cm 1cm 1.7cm 1.5cm}, clip, width=.9\linewidth]{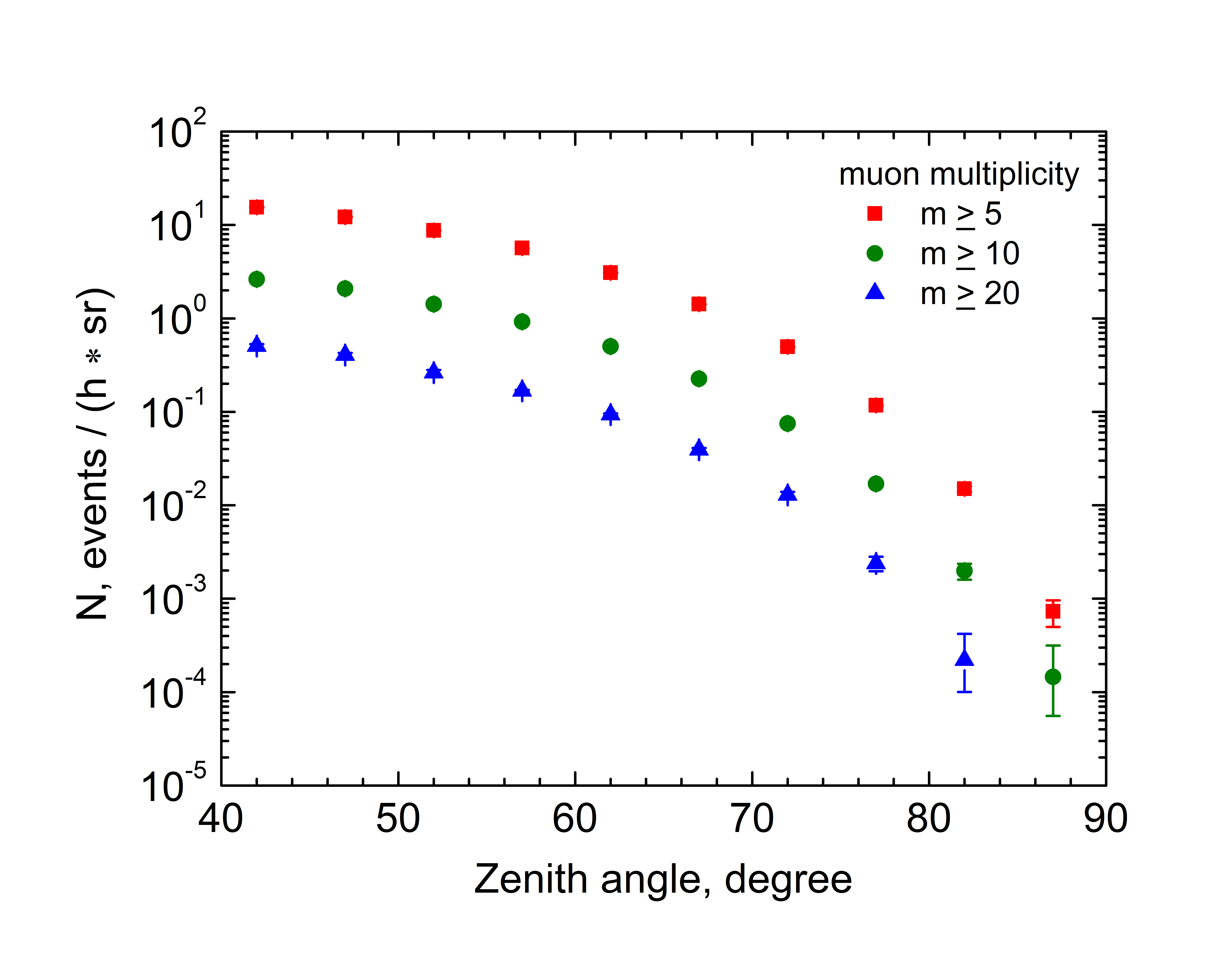}
 \end{minipage}
\caption{\label{fig:6} The dependencies of the intensity of muon bundles detected by the NEVOD-DECOR on the multiplicity (top) and on the zenith angle (bottom).}
\end{figure}

The main characteristics of muon bundles detected at the NEVOD-DECOR are shown in Fig.~\ref{fig:6}: the dependences of intensity on the multiplicity for different intervals of zenith angles (top) and the dependences of intensity on the zenith angle for different multiplicities (bottom). As can be seen from the figure, in a single experiment it was possible to cover a range of more than five orders of magnitude in the intensity of muon bundles, which corresponds to a range of more than three orders of magnitude in the primary CR energy.

\section{Data analysis (Local Muon Density Spectra)}

To analyze the muon bundle data of the NEVOD-DECOR experiment (and compare them with calculations), the original approach to studying EAS muon component, the method of local muon density spectra (LMDS)~\cite{Bogdanov2010,Bogdanov2018,Kokoulin2021}, was developed. To some extent, it is a successor of the shower density spectrum~\cite{Grieder2010} which was widely used to study air showers after their discovery and, in fact, reflects the power-law behavior of the primary CR energy spectrum.

\subsection{Observed LMDS}

If the typical transverse dimensions of the EAS muon component significantly exceed the size of the setup (as in our case), the detector can be considered as a point-like. Then, in the first approximation, the local muon density \textit{D} at the observation point can be estimated as the ratio of the number of muons m hitting the detector (the multiplicity of muons in the bundle) to the effective area $S_\text{det}(\theta,\phi)$ in a certain direction. The distribution of events by D, in fact, forms the LMDS.

A detailed analysis of the distributions of muon bundles detected at the NEVOD-DECOR by the multiplicity and zenith angle (Fig.~\ref{fig:6}) showed that they are rather well described by a phenomenological model of the LMDS of a very simple form:
\begin{equation}
dF_\text{0}/dD = CD^{-(\beta+1)}cos^{\alpha}\theta,
\label{eq:1}
\end{equation}
with parameters $C = 7 \times 10^{-4}, \alpha = 4.7, \beta = 2.1$, which were obtained using the maximum likelihood method.

Experimental estimates of the differential spectrum of the local muon density in a detector-independent form were calculated as:
\begin{eqnarray}
(dF/dD)_\text{obs} &&= [\Delta N_\text{obs}(\Delta D, \Delta \theta)] / [\Delta N_\text{exp}(\Delta D, \Delta \theta)] \nonumber \\ 
&&\times dF_\text{0}/dD.
\label{eq:2}
\end{eqnarray}

The matrix of detected events $\Delta N_\text{obs}(\Delta D, \Delta \theta)$ was constructed by sorting them by zenith angle (with a step of 5$^{\circ}$) and by the estimate of the local muon density $D_\text{est}$ (with a constant logarithmic step):
\begin{equation}
D_\text{est} = (m-\beta)/S_\text{det}(\theta, \phi),
\label{eq:3}
\end{equation}
taking into account its bias due to the steeply falling power-law density spectrum and to the Poisson fluctuations in the number of particles hitting the detector~\cite{Bogdanov2010}. For further physical analysis, those matrix cells, where the number of events was at least 10 – 20 and the local muon density value $D_\text{est}$ was less or equal to 2.0 m$^{-2}$, were used (see Table~\ref{tab:1} of the Appendix).

The expected matrix $\Delta N_\text{exp}(\Delta D, \Delta \theta)$ was calculated using the Monte Carlo technique. Artificial events were sampled in terms of muon density and arrival direction according to the LMDS reference model (\ref{eq:1}). Poisson fluctuations in the number of muons hitting the SMs, the arrangement of the DECOR supermodules in the experiment, the triggering efficiency of the streamer tube planes, as well as all the conditions of the hardware, software and operator event selection have been taken into account. In particular, the probability of track masking (loss) due to the finite resolution of the DECOR detector was considered. Events with artificial muon bundles were sorted by the local density $\Delta D$ and zenith angle $\Delta \theta$, as it is done in the experimental data processing. But the number of such events is two orders of magnitude greater than in the experiment.

The LMDS estimates for each matrix cell were assigned to certain values ($D^{*}$, $\theta^{*}$), the choice of which was optimized in terms of the robustness of reconstruction procedure to variations in the $\alpha$ and $\beta$ parameters of the LMDS reference model within reasonable limits. The mean logarithmic values of the local muon density were used as $D^{*}$, and the rounded value of the zenith angle, providing a cosine value close to the mean logarithmic value, was used as $\theta^{*}$ (in degrees).

\subsection{Expected LMDS}

\begin{figure}[h!]
 \begin{minipage}[h]{0.475\textwidth}
  \includegraphics[trim={2.8cm 1.5cm 1.5cm 3.3cm}, clip, width=.75\linewidth]{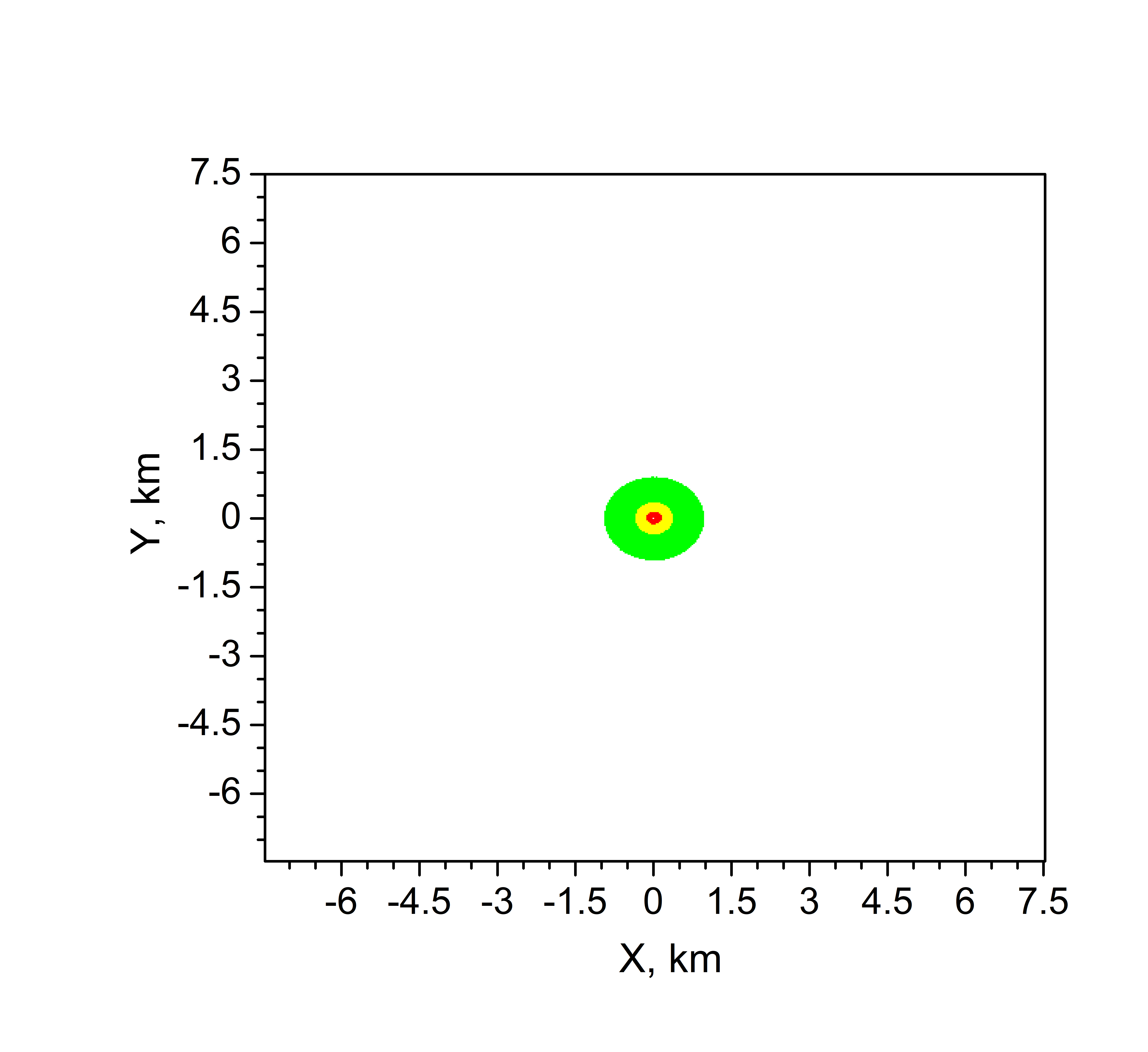}
 \end{minipage}
 \begin{minipage}[h]{0.475\textwidth}
  \includegraphics[trim={2.8cm 1.5cm 1.5cm 3cm}, clip, width=.75\linewidth]{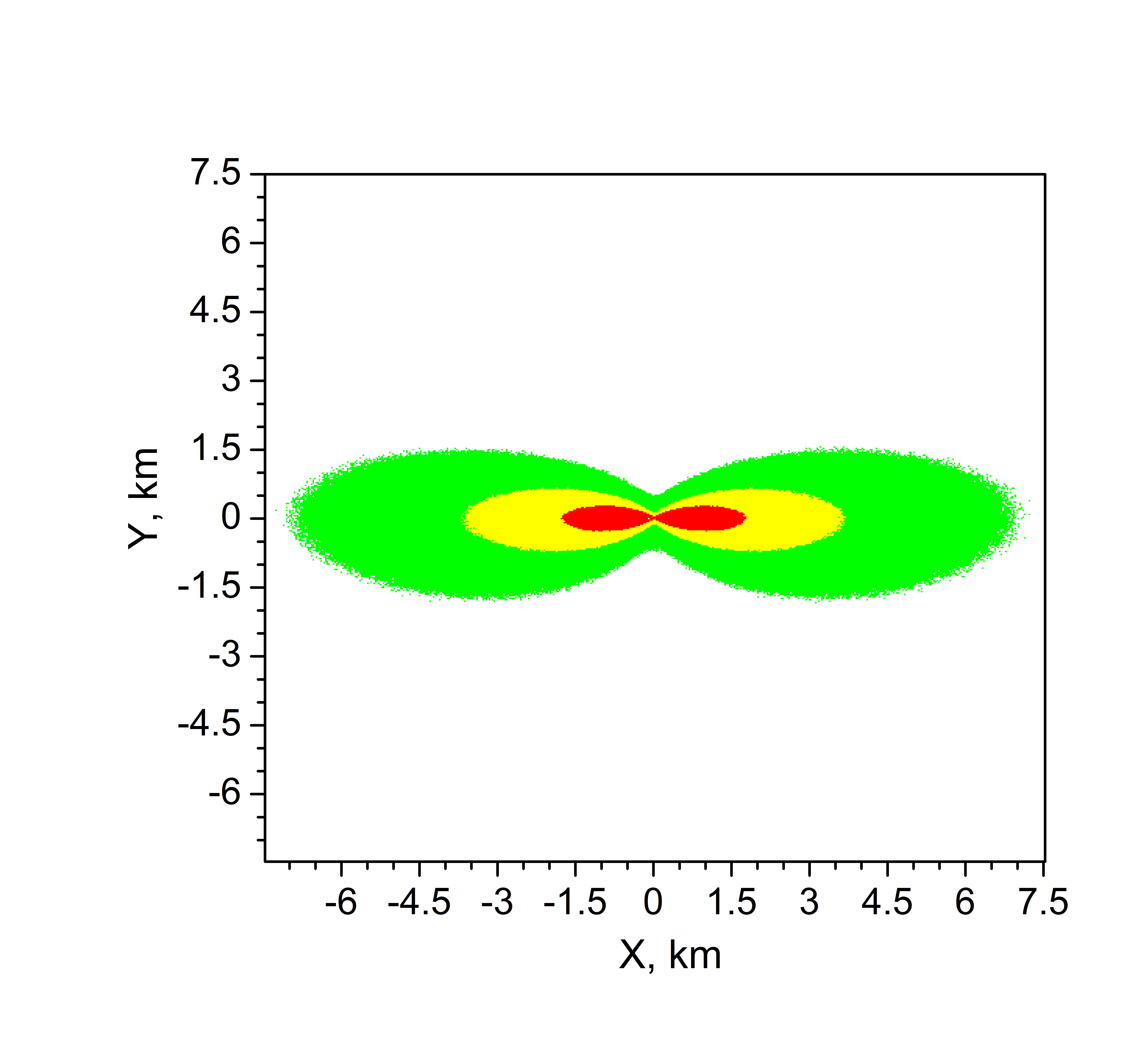}
 \end{minipage}
\caption{\label{fig:7} The transverse cross section of a simulated EAS from proton with energy $E_\text{0} = 10^{18}$ eV arriving at zenith angle $\theta = 42^{\circ}$ (top) and $\theta = 82^{\circ}$ (bottom) taking into account the Earth's magnetic field. The muon content in each shaded area is about 30$\%$ of the total number.}
\end{figure}

\begin{figure}[h!]
\includegraphics[trim={1.8cm 1.5cm 2.0cm 2.3cm}, clip, width=.9\linewidth]{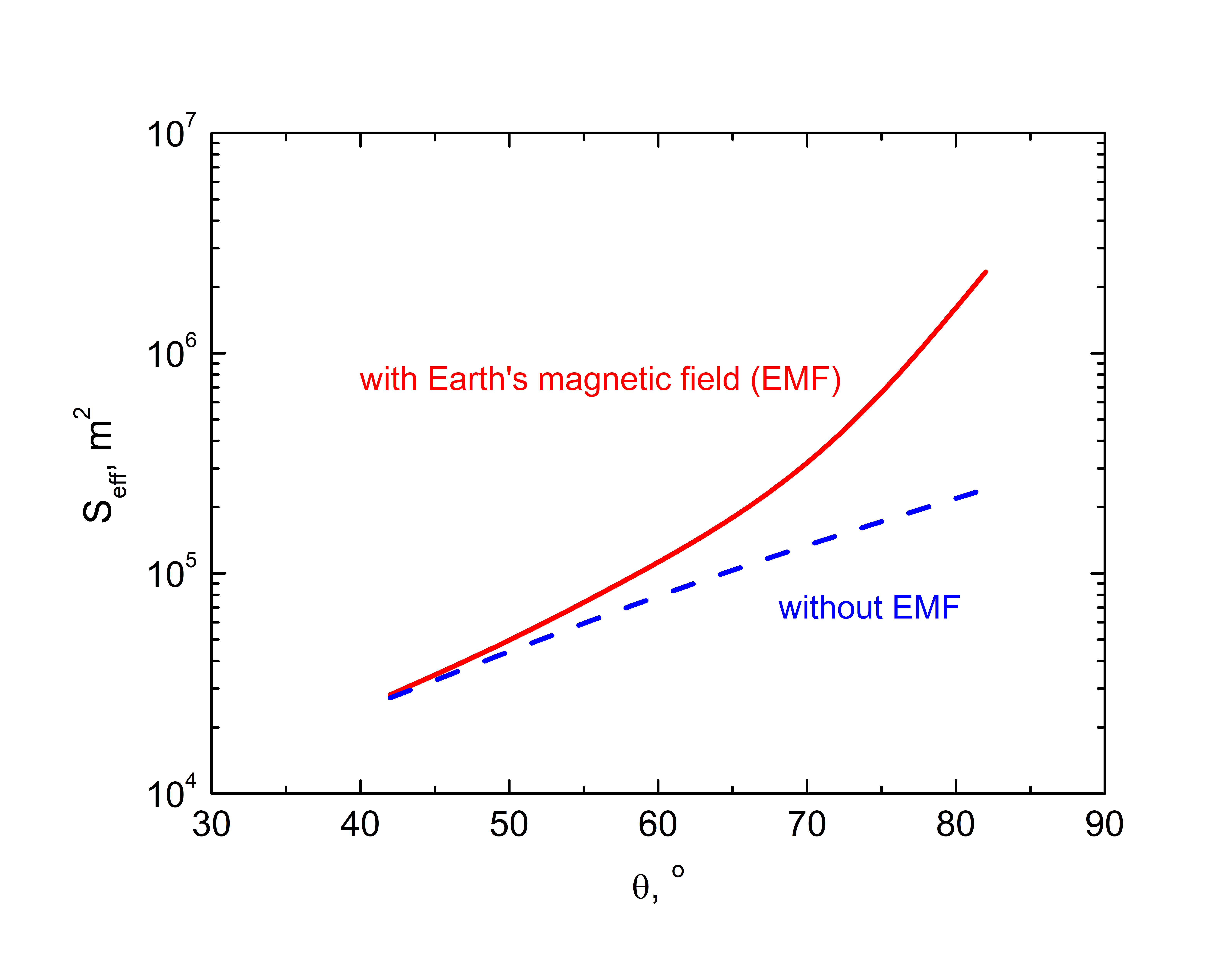}
\caption{\label{fig:8} The dependence of the effective event collection area on the zenith angle for the LMDS method (air showers are initiated by protons with energy $E_\text{0} = 10^{18}$ eV).}
\end{figure}

The local muon density spectrum is formed by air showers at random distances from the axis. In order to calculate the expected integral intensity of events $F(\geq D)$ in which the local muon density exceeds a certain value \textit{D}, it is necessary to integrate over the EAS cross section. In this case, the event collection area is determined not by the sizes of the experimental setup (which are small), but by the dimensions of the shower cross section in the muon component. As the zenith angle increases, the characteristic transverse dimensions of the shower in muons (Fig.~\ref{fig:7}) and the effective event collection area (Fig.~\ref{fig:8}) increase. For near-horizontal showers, its value reaches the order of several square kilometers, which is sufficient for studying the flux of primary particles with energies of $\sim 10^{18}$ eV and higher.

According to~\cite{Bogdanov2010}, for a given direction, the integral LMDS can be represented as:
\begin{equation}
F(\geq D) = \int N(\geq E_\text{0} (\textbf{r}, D)) dS,
\label{eq:4}
\end{equation}
where $N(\geq E_\text{0} (\textbf{r}, D))$ is the integral energy spectrum of the primary CRs, \textbf{r} is the distance between the observation point (detector) and the air shower axis. The minimum energy $E_\text{0}$ is determined by solving the equation:
\begin{equation}
\rho(E_\text{0},\textbf{r}) = D,
\label{eq:5}
\end{equation}
where $\rho(E_\text{0},\textbf{r})$ is the muon lateral distribution function (LDF).

Within some approximations (the power-law CR spectrum and the scaling behavior of the muon LDF shape with a change of the primary particle energy), LMDS also has a power-law shape with a slope that is greater than the slope of the primary CR spectrum.

From Eqs. (\ref{eq:4}) and (\ref{eq:5}), it is easy to obtain a formula for calculating the differential LMDS:
\begin{equation}
dF/dD = \int (dN/dE_\text{0})/[d\rho(E_\text{0},\textbf{r})/dE_\text{0}]dS.
\label{eq:6}
\end{equation}

As follows from Eqs. (\ref{eq:4}) – (\ref{eq:6}), the expected local muon density spectrum is a convolution of the model of the primary CR energy spectrum and the muon LDFs. In turn, the shape of the muon LDF depends on the choice of the hadronic interaction model and the type (atomic weight) of the primary CR nucleus.

\begin{figure}[h]
\includegraphics[trim={1.7cm 1.5cm 2.0cm 1.9cm}, clip, width=.9\linewidth]{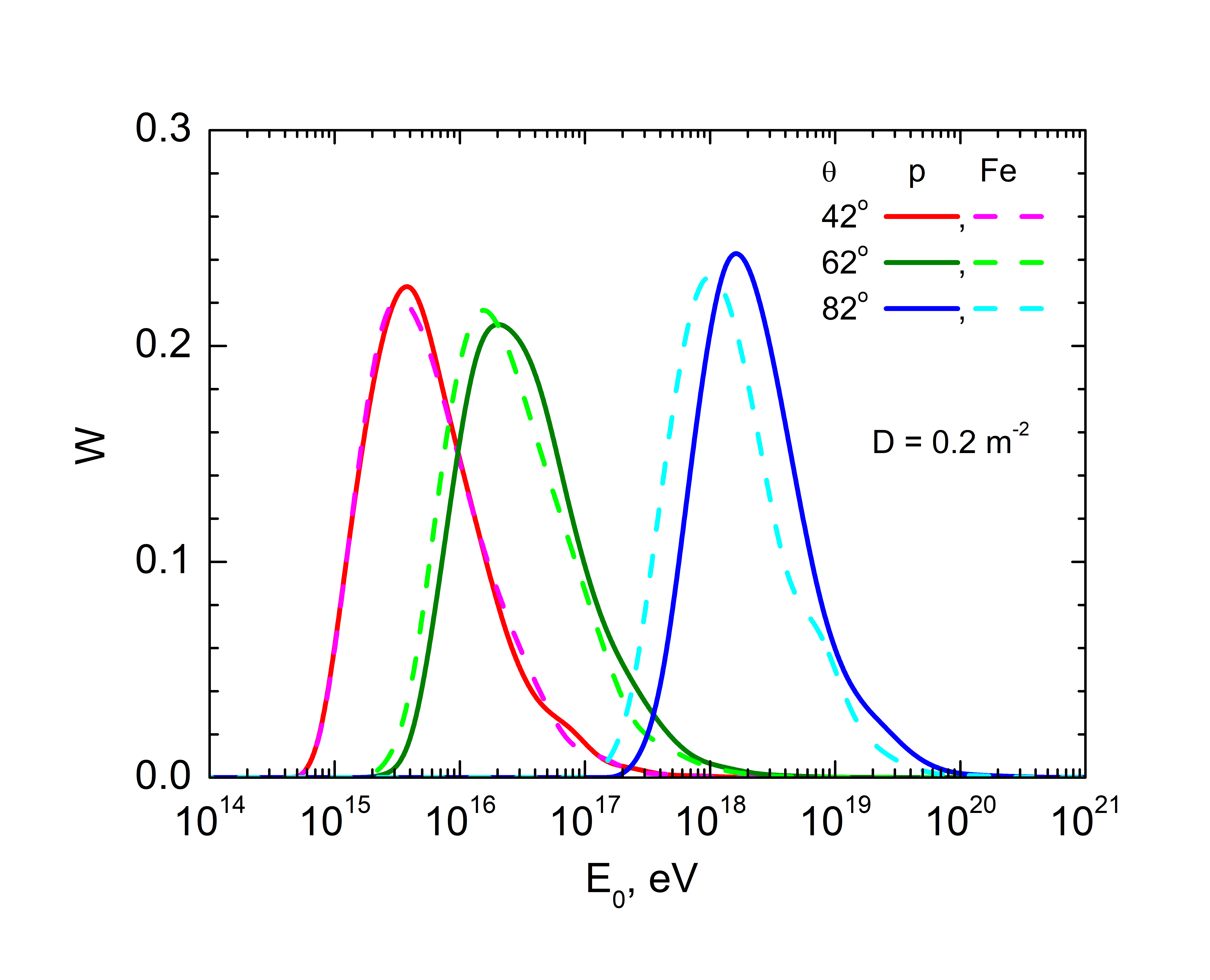}
\caption{\label{fig:9} The energy distributions of primary CRs contributing to events with a certain local muon density ($D = 0.2$ m$^{-2}$) at different zenith angles.}
\end{figure}

It should be noted that air showers with different energies contribute to events with a certain local muon density. However, estimates based on the results of EAS simulation using the CORSIKA program~\cite{Heck1998} showed, that the effective energy range of primary particles is relatively narrow due to the rapid decrease in the CR intensity with increasing energy. As an illustration, Fig.~\ref{fig:9} shows for three different zenith angles the energy distributions of primary particles contributing to events with a fixed local muon density $D \approx 0.2$ m$^{-2}$, which corresponds to approximately 5 - 7 muons passed through the DECOR detector. The width of distributions is $\Delta \text{log}_{10}E_\text{0} \approx 0.4$.

\begin{figure*}[ht]
\includegraphics[trim={1cm 1cm 1cm 1cm}, clip, width=.95\linewidth]{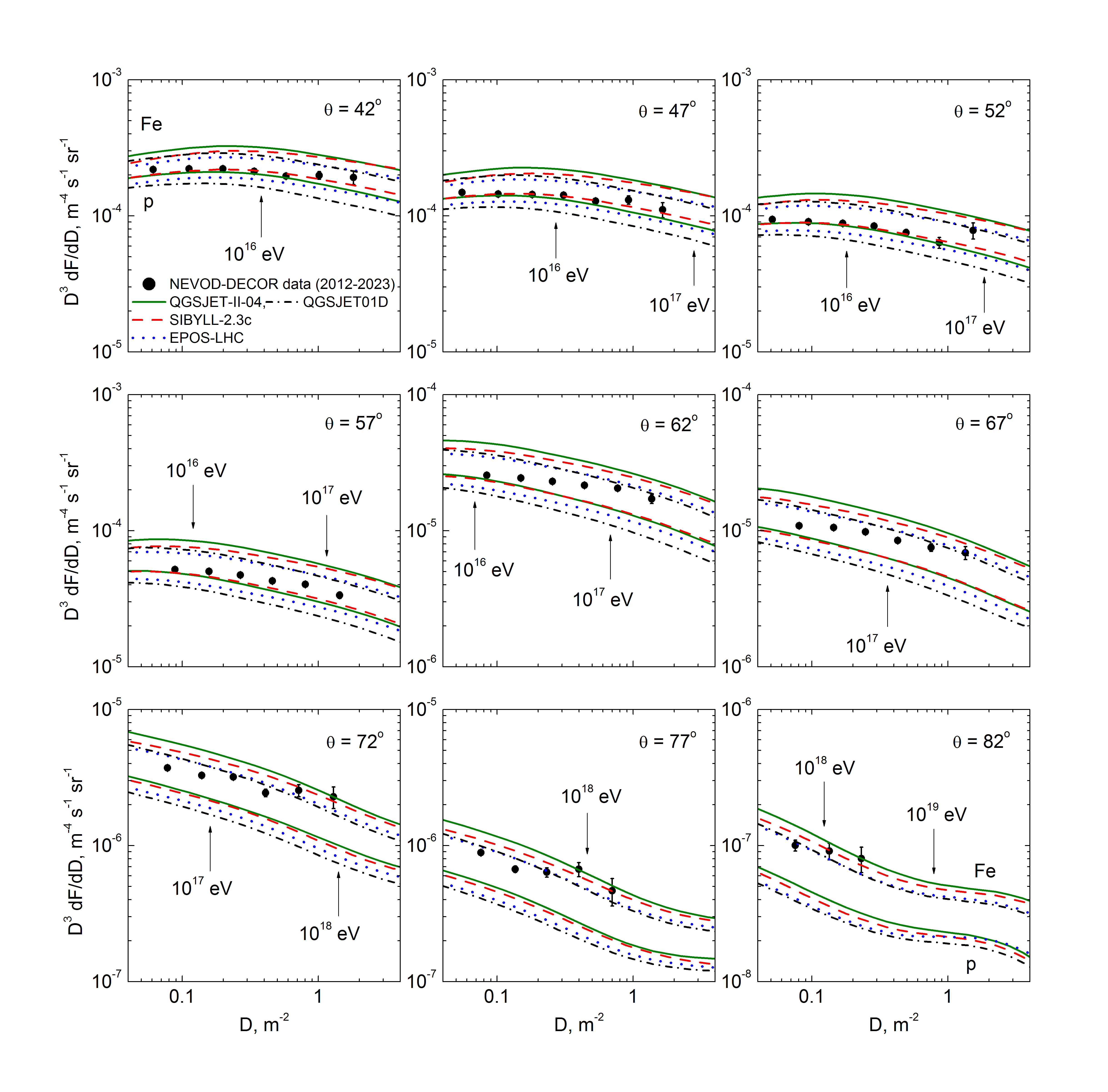}
\caption{\label{fig:10}  The differential local muon density spectra for nine zenith angle intervals. The symbols show the NEVOD-DECOR data. The curves are the expected LMDS for four hadronic interaction models (EPOS-LHC, QGSJET-II-04, SIBYLL-2.3c, QGSJET01D) and two primary nuclei (p, Fe). The arrows show the typical energies of the primary particles.}
\end{figure*}

It is important to emphasize, that for the same local muon density, different zenith angles correspond to significantly (by orders of magnitude) different effective energies of primary particles. With increasing zenith angle, muons pass greater distances in the atmosphere, and their spread in bundles increases (therefore, the density decreases). That is due to transverse momenta during the production and decay of parent hadrons, deviation in the Earth's magnetic field (EMF) and multiple scattering. Thus, measuring the LMDS over a wide range of zenith angles make it possible to study the characteristics of the flux and interactions of CRs in a wide energy range using a relatively small setup.

The muon LDFs were calculated based on the EAS simulations in the CORSIKA program (version 7.6900) with the CURVED option, which allows taking into account the curvature of the atmosphere at zenith angles greater than 70$^{\circ}$. Four models of high-energy hadronic interactions EPOS-LHC, QGSJET-II-04, SIBYLL-2.3c (post-LHC), QGSJET01D (pre-LHC) and two extreme assumptions about the mass composition of primary CRs (protons and iron nuclei) were used. Hadronic interactions with energies below 80 GeV were treated using the FLUKA2011. All interactions of electrons and photons in the atmosphere were treated with the EGS4 code explicitly, and the Landau-Pomeranchuk-Migdal effect was activated. Full Monte Carlo simulation was performed for the QGSJET-II-04 model. After choosing the optimal (speed/accuracy) parameters for the particle thin sampling mechanism implemented in the CORSIKA, the THIN option was used to reduce the computing time for other models.

Moreover, simulation was carried out for three recently released high-energy hadronic interaction models (with default parameters): EPOS LHC-R, QGSJET-III-01 and SIBYLL-2.3e, included in the actual version 7.8010 of the CORSIKA program. In this case, low-energy hadronic interactions were treated using the FLUKA2024 (INFN) model.

The simulation was performed for primary particles with fixed energies $E_0$ in the range from $10^{13}$ to $10^{20}$ eV (step by logarithm of energy $= 0.5)$ and with zenith angles in the range from 42$^{\circ}$ to 87$^{\circ}$ (step by angle $\Delta\theta = 5^{\circ}$). The low energy cut-off was 1 GeV for hadrons, muons, electrons and photons. The standard U.S. atmosphere was used in the simulation. The observation level was taken to be 160 m s.l. The EMF was taken into account in the simulation using average values of the horizontal and vertical components for the geographic location of the NEVOD-DECOR. The simulation statistics ranged from 1 million events for energy $E_\text{0} = 10^{13}$ eV to 150 events for $E_\text{0} = 10^{20}$ eV.

For given primary particle type, energy and zenith angle, the two-dimensional LDFs of muons with energies greater than 2 GeV, which is the threshold in the experiment, were obtained using the simulated showers. The need to calculate two-dimensional LDFs is due to the violation of their axial symmetry under the influence of the Earth's magnetic field~\cite{Bogdanov2007}, especially at large zenith angles (see Fig.~\ref{fig:7}).

As the basic model of the differential energy spectrum of the primary CRs (all-particle), a piecewise power-law approximation N$\&$D~\cite{Bogdanov2018} of data of various experiments from the Particle Data Group review~\cite{Patrignani2016} was used:
\begin{eqnarray}
&&dN_\text{1}/dE_\text{0} = 4.2 \times 10^4 \times (E_\text{0},\text{GeV})^{-2.7}, E_\text{0} \leq E_\text{1}, \nonumber\\ 
&&dN_\text{2}/dE_\text{0} = dN_\text{1}/dE_\text{0} \times (E_\text{0}/E_\text{1})^{-0.35}, E_\text{1} < E_\text{0} \leq E_\text{2}, \nonumber\\
&&dN_\text{3}/dE_\text{0} = dN_\text{2}/dE_\text{0} \times (E_\text{0}/E_\text{2})^{-0.25}, E_\text{2} < E_\text{0} \leq E_\text{3}, \nonumber\\
&&dN_\text{4}/dE_\text{0} = dN_\text{3}/dE_\text{0} \times (E_\text{0}/E_\text{3})^{0.6}, E_\text{3} < E_\text{0} \leq E_\text{4}, \nonumber\\
&&dN_\text{5}/dE_\text{0} = dN_\text{4}/dE_\text{0} \times (E_\text{0}/E_\text{4})^{-1.5}, E_\text{0} > E_\text{4}, 
\label{eq:7}
\end{eqnarray}
where dimension is $\left[\text{m}^{-2}\text{s}^{-1}\text{sr}^{-1}\text{GeV}^{-1} \right]$, $E_1 = 4 \times 10^6$, $E_2 = 2 \times 10^8$, $E_3 = 4 \times 10^9$, $E_4 = 4 \times 10^{10}$ GeV.

\subsection{Comparison of observed and expected LMDS}

A comparison of the experimental and expected differential LMDS for nine zenith angle intervals is shown in Fig.~\ref{fig:10}. The NEVOD-DECOR data are marked with circles, and the numerical values are given in Table~\ref{tab:2} of the Appendix. The solid, dashed, dotted, and dash-dotted curves represent the results of CORSIKA based calculations for different high-energy hadronic interaction models: EPOS-LHC, QGSJET-II-04, SIBYLL-2.3c, and QGSJET01D, respectively. The lower curves correspond to the predictions for primary protons, the upper ones correspond to the predictions for iron nuclei. The arrows in the figure indicate the effective (mean logarithmic) energies of primary CRs (protons), which give the main contribution to the events selected by means of the local muon density. The ratio of the theoretical LMDS for the post- and pre-LHC models is approximately 1.15 – 1.35 in the region of interest of local densities, and for the showers initiated by iron nuclei and protons it varies from 1.5 to 2.5 (depending on the zenith angle). It is important that replacing the all-particle spectrum model N$\&$D (\ref{eq:7}) with some other one will change the expected LMDS curves.

The shift of the experimental points relative to the theoretical curves with an increase in the zenith angle from $\theta = 42^{\circ}$ to $82^{\circ}$ and, accordingly, the energy of the primary particles from approximately $2 \times 10^{15}$ to $3 \times 10^{18}$ eV, in principle, could be interpreted as an evolution of the CR mass composition from light nuclei to heavy ones, if we remain within the existing hadronic interaction models and the specified all-particle energy spectrum model.

\begin{figure}[h!]
 \begin{minipage}[h]{0.475\textwidth}
  \includegraphics[trim={1.5cm 1.5cm 2.0cm 2cm}, clip, width=.85\linewidth]{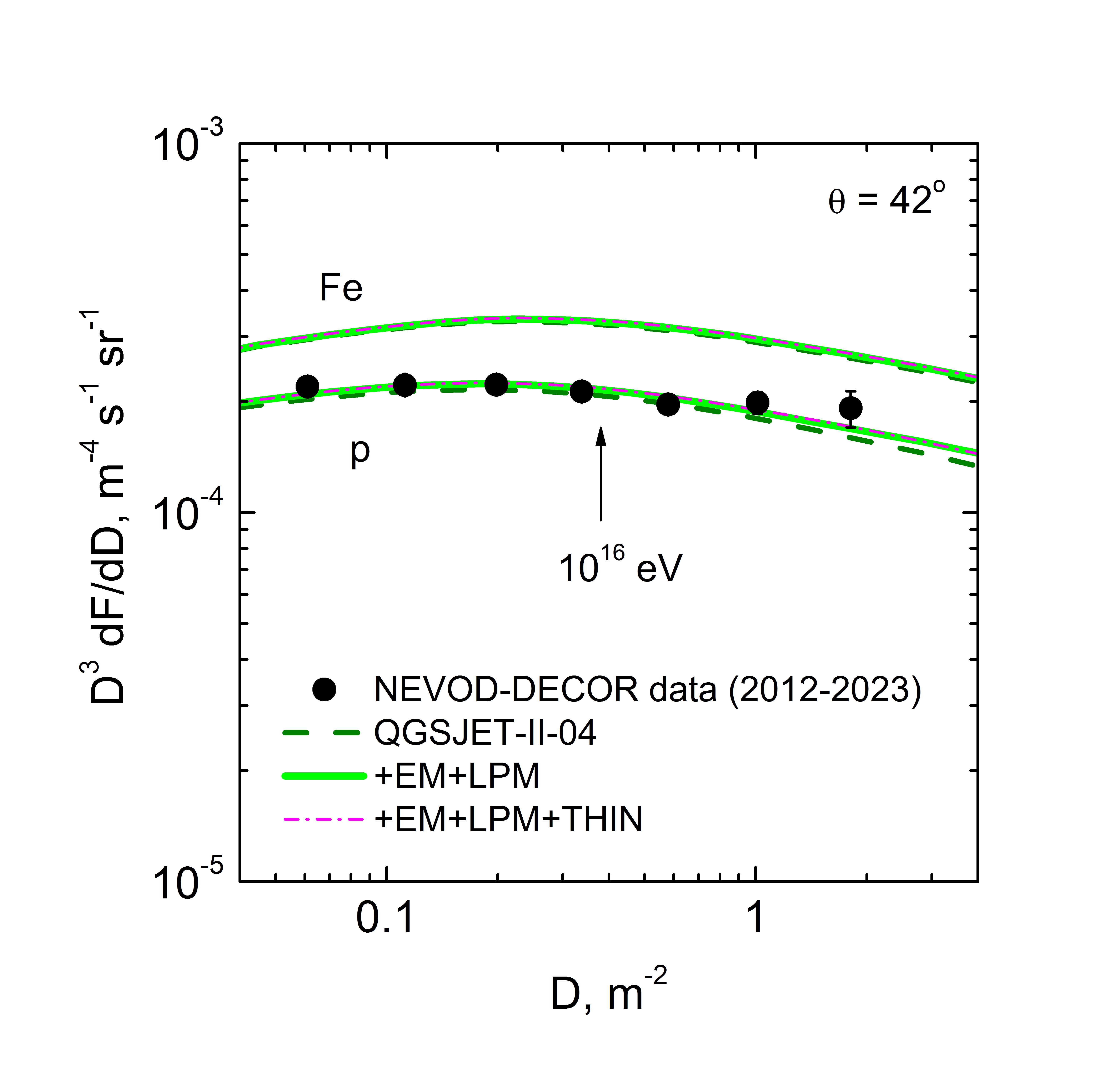}
 \end{minipage}
 \begin{minipage}[h]{0.475\textwidth}
  \includegraphics[trim={1.5cm 1.5cm 2.0cm 2cm}, clip, width=.85\linewidth]{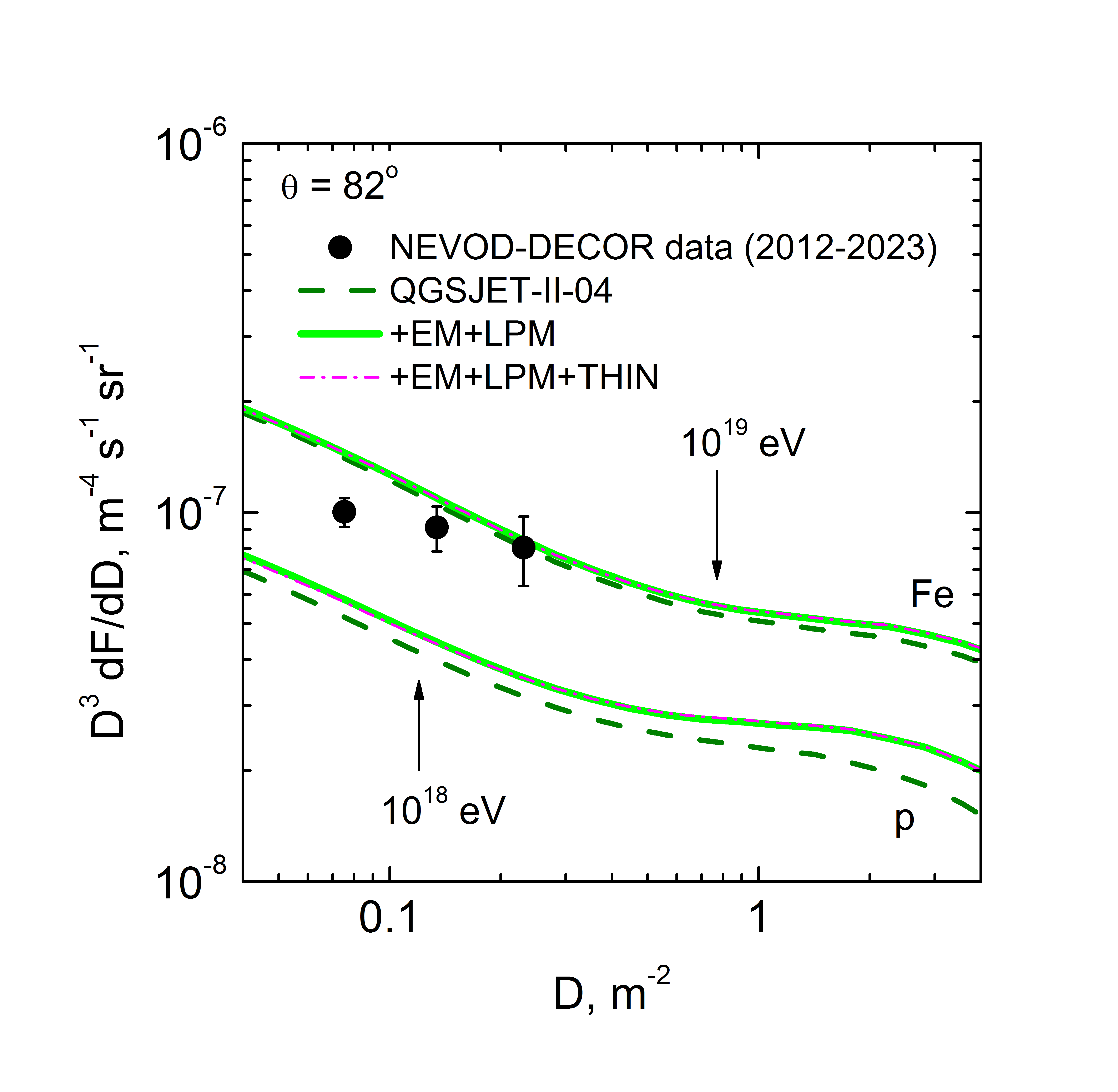}
 \end{minipage}
\caption{\label{fig:11}  The differential local muon density spectra for two zenith angle intervals. The symbols show the NEVOD-DECOR data. The curves show the expected LMDS for one hadronic interaction model QGSJET-II-04 and two primary nuclei (p, Fe) taking into account the electron-photon component of the air showers (+EM+LPM, solid) and without it (dashed), the dashed-dotted lines (+EM+LPM+THIN, almost overlap with solid ones) show the expected LMDS obtained using the thinning mechanism in the EAS simulation.}
\end{figure}

The influence of the EAS electron-photon component on the LMDS is quite noticeable, as illustrated in Fig.~\ref{fig:11}. The dots in the figure show the LMDS obtained from the NEVOD-DECOR data for two intervals of zenith angles ($\theta =$ 40 – 45$^{\circ}$, top) and ($\theta =$ 80 – 85$^{\circ}$, bottom), and the lines show the theoretical LMDS (for the QGSJET-II-04 model) taking into account the electron-photon component of the air showers (solid) and without it (dashed). The maximum difference between the latter in the region of interest (i.e., the experimental LMDS) for EAS from primary protons (lower curves) is about 7$\%$ at moderate zenith angles and about 13$\%$ at large zenith angles. For EAS initiated by iron nuclei, the difference of the LMDS (upper curves) is much smaller. The main contribution to the generation of muons (through hadrons, which can then decay) is given by the photonuclear interaction, while the contribution of the process of muon pair production by a photon is negligibly small. In the same figure, the dashed-dotted lines show the calculated LMDS obtained as a result of the EAS simulation using the particle thinning mechanism (THIN option). Their difference from the LMDS obtained as a result of the complete simulation of all air shower particles does not exceed 1.5$\%$.

\begin{figure*}[ht!]
\includegraphics[trim={1cm 1cm 1cm 1cm}, clip, width=0.75\linewidth]{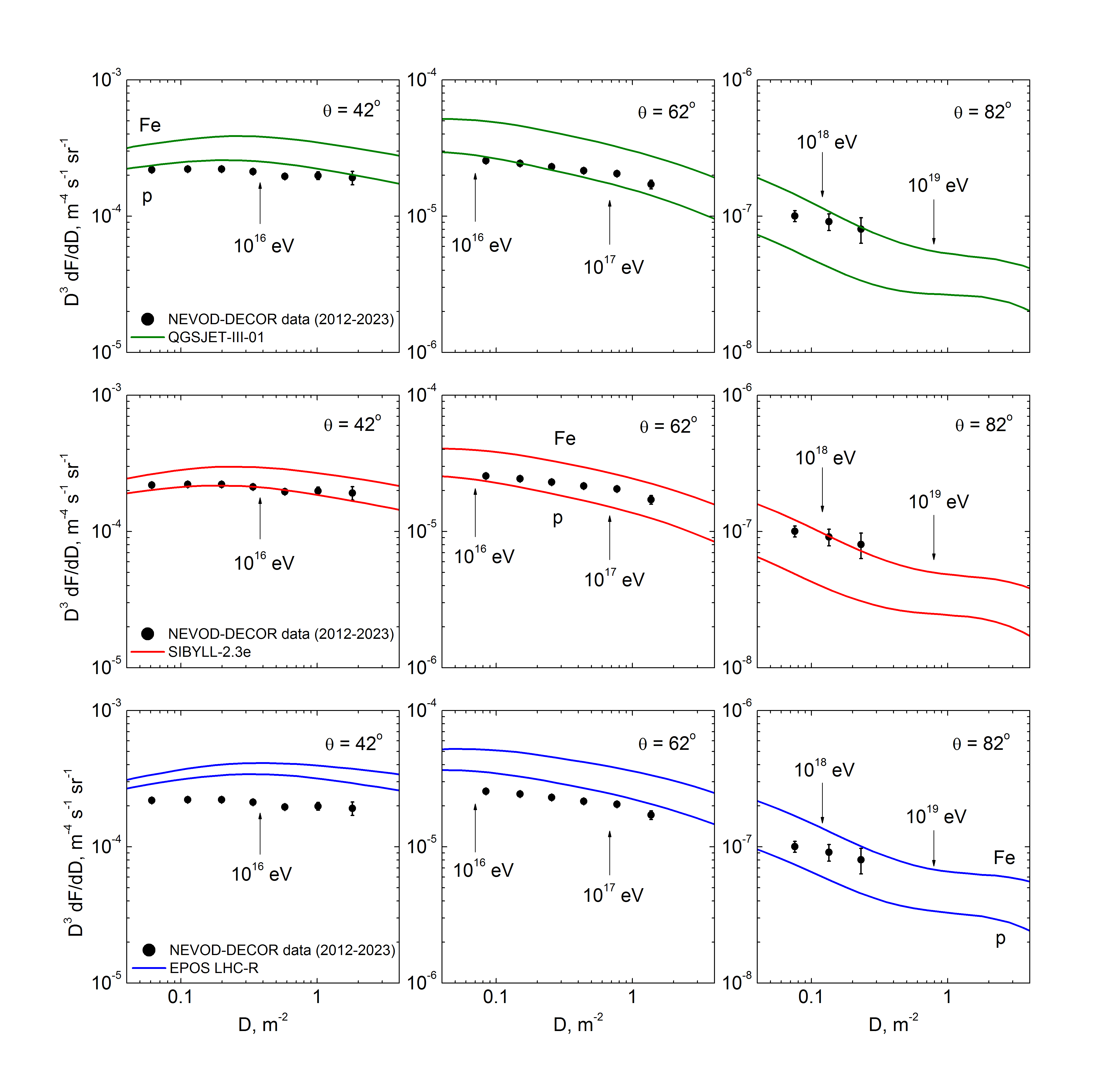}
\caption{\label{fig:12}  The differential local muon density spectra for three zenith angle intervals. The symbols show the NEVOD-DECOR data. The curves show the expected LMDS for three recently released hadronic interaction models (QGSJET-III-01, SIBYLL-2.3e, EPOS LHC-R) and two primary nuclei (p, Fe).}
\end{figure*}

Of undoubted interest is the comparison of the NEVOD-DECOR experiment data on muon bundles and calculations using the recently released models of high-energy hadronic interactions: EPOS LHC-R, QGSJET-III-01 and SIBYLL-2.3e. Fig.~\ref{fig:12} shows the LMDS for three intervals of zenith angles: $\theta$ = 40 – 45$^{\circ}$ (left), $\theta =$ 60 – 65$^{\circ}$ (center) and $\theta =$ 80 – 85$^{\circ}$ (right). The dots represent the NEVOD-DECOR data, the curves show the expected LMDS for three models: QGSJET-III-01 (three top plots), SIBYLL-2.3e (middle ones), EPOS LHC-R (bottom ones). The upper curve for each model corresponds to the EAS from primary iron nuclei, and the lower one is for the EAS from protons (as in Fig.~\ref{fig:10}).

It is obvious that, in contrast to other models of high-energy hadronic interactions, the expected LMDS for the new EPOS LHC-R model (with default options) strongly contradict the experimental LMDS at moderate zenith angles ($\theta = 42^{\circ}, 62^{\circ}$). Note that the calculations use an approximation of the PCR energy spectrum (\ref{eq:7}) based on the data of the vast majority of experiments on measuring the EAS electron-photon component using surface detectors, which also overlap with the data of direct observations (on satellites and balloons). To eliminate the contradiction between the expected (within the EPOS LHC-R model) LMDS and the NEVOD-DECOR data, it is required to reduce the absolute intensity of the PCR flux at energies of $10^{15}$ – $10^{16}$ eV by about 1.5 times under the assumption of a light mass composition of primaries (p, He) in this energy range. Such low estimates of the PCR flux intensity are obtained, for example, from the data of the High-Elevation Auger Telescopes (HEAT), which record events with dominant Cherenkov radiation of air showers, or from the data of the completed Chicago Air Shower Array (CASA) experiment. The discrepancy between the results of calculations using the EPOS LHC-R model and the NEVOD-DECOR experimental data on muon bundles decreases with the growth of zenith angle, and at large zenith angles, which correspond to primary particle energies of $\sim 10^{18}$ eV, no contradiction is observed. The expected LMDS for the new QGSJET-III-01 model are in poor agreement with the NEVOD-DECOR data at moderate zenith angles. At large zenith angles, the QGSJET-III-01 model does not lead to noticeable changes compared to the previous version (QGSJET-II-04). For the new SIBYLL-2.3e model (compared to the previous version, SIBYLL-2.3c), no drastic changes in the expected LMDS, that could affect the interpretation of the NEVOD-DECOR data, are visible.

\begin{figure*}[ht]
\includegraphics[trim={1.5cm 1.5cm 2.5cm 2.5cm}, clip, width=.9\linewidth]{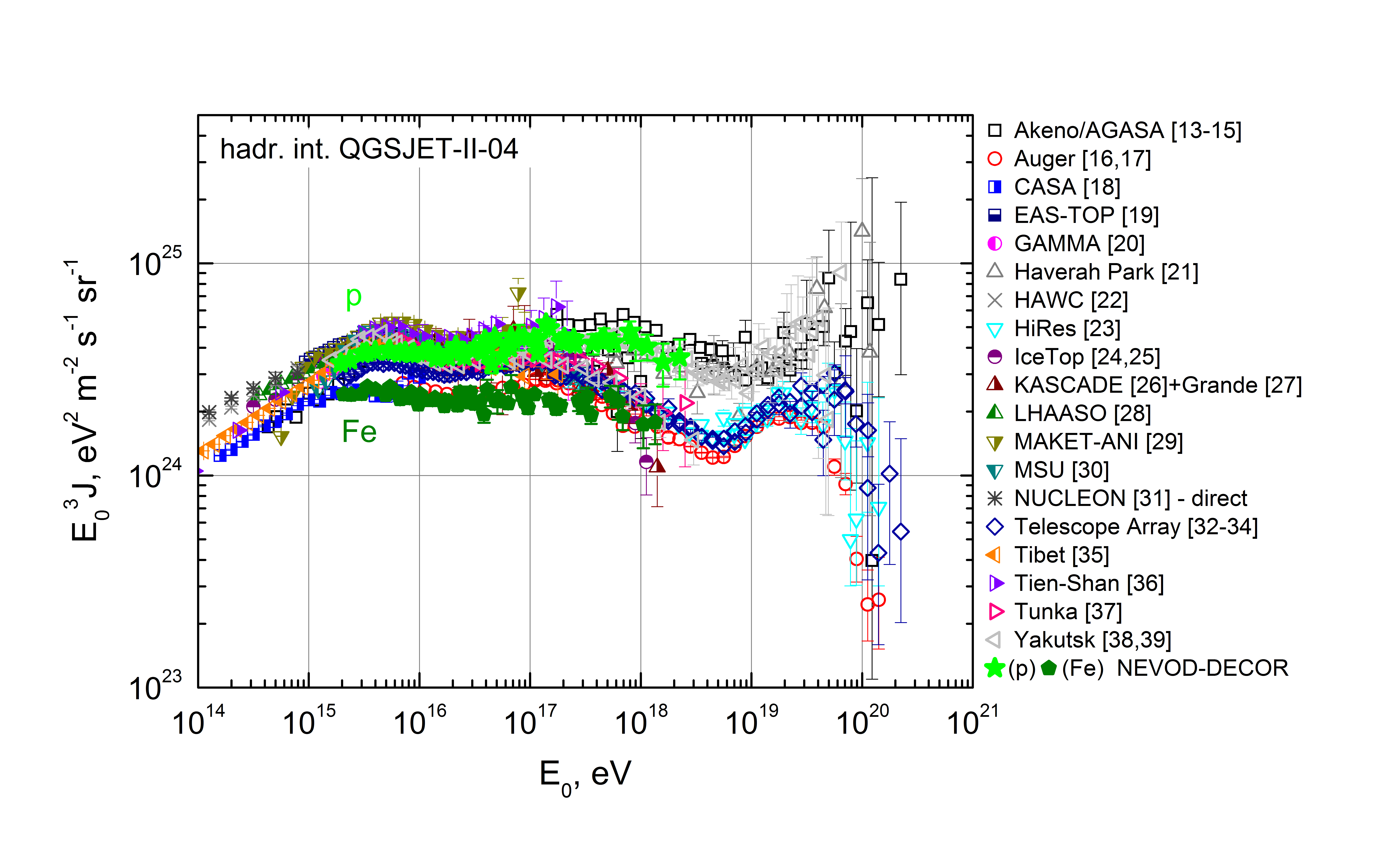}
\caption{\label{fig:13}  The differential energy spectra of the primary CRs obtained from the intensity of muon bundles detected at the NEVOD-DECOR setup, under the assumption that all primary particles are protons (stars) or iron nuclei (pentagons). The hadronic interaction model is QGSJET-II-04. Data from a set of experiments obtained by analyzing the EAS electron-photon component are given for comparison.}
\end{figure*}

\section{Estimates of the primary cosmic ray energy spectrum}

Muon bundles are part of the muon component of air showers, so their intensity can be directly recalculated to the intensity of the primary CR flux, but under certain assumptions about the mass composition and the hadronic interaction model.

Reconstruction of the differential CR energy spectrum from the experimental LMDS (for each zenith angle interval) was carried out according to the following formula:
\begin{eqnarray}
dJ/dE_\text{0} &&= \left[ (dF/dD)_\text{obs}/(dF/dD)_\text{sim} \right] \nonumber \\ 
&&\times (dN/dE_\text{0})_\text{mod}, 
\label{eq:8}
\end{eqnarray}
where $(dN/dE_\text{0})_\text{mod}$ is the model of the CR spectrum, $(dF/dD)_\text{obs}$ is the experimental LMDS, $(dF/dD)_\text{sim}$ is the theoretical LMDS (convolution of the muon LDF with the spectrum model).

Estimates of $E_\text{0}$ were obtained from the calculated distributions of primary particles by energies that give the main contribution to events with a certain muon density $D_\text{0}$ (see Fig.~\ref{fig:9}). Besides, the mean logarithmic values were used, as the most optimal for quasi power-law spectra~\cite{Kokoulin1988} (as opposed to mean or median ones). Then, the estimation of the CR flux intensity at point $E_\text{0}$ based on the measured distribution at point $D_\text{0}$ is the least sensitive to the variation of the spectrum slope.

\begin{figure*}[ht]
 \begin{minipage}[h]{0.495\textwidth}
  \includegraphics[trim={1.5cm 1.5cm 2cm 2,2cm}, clip, width=.975\linewidth]{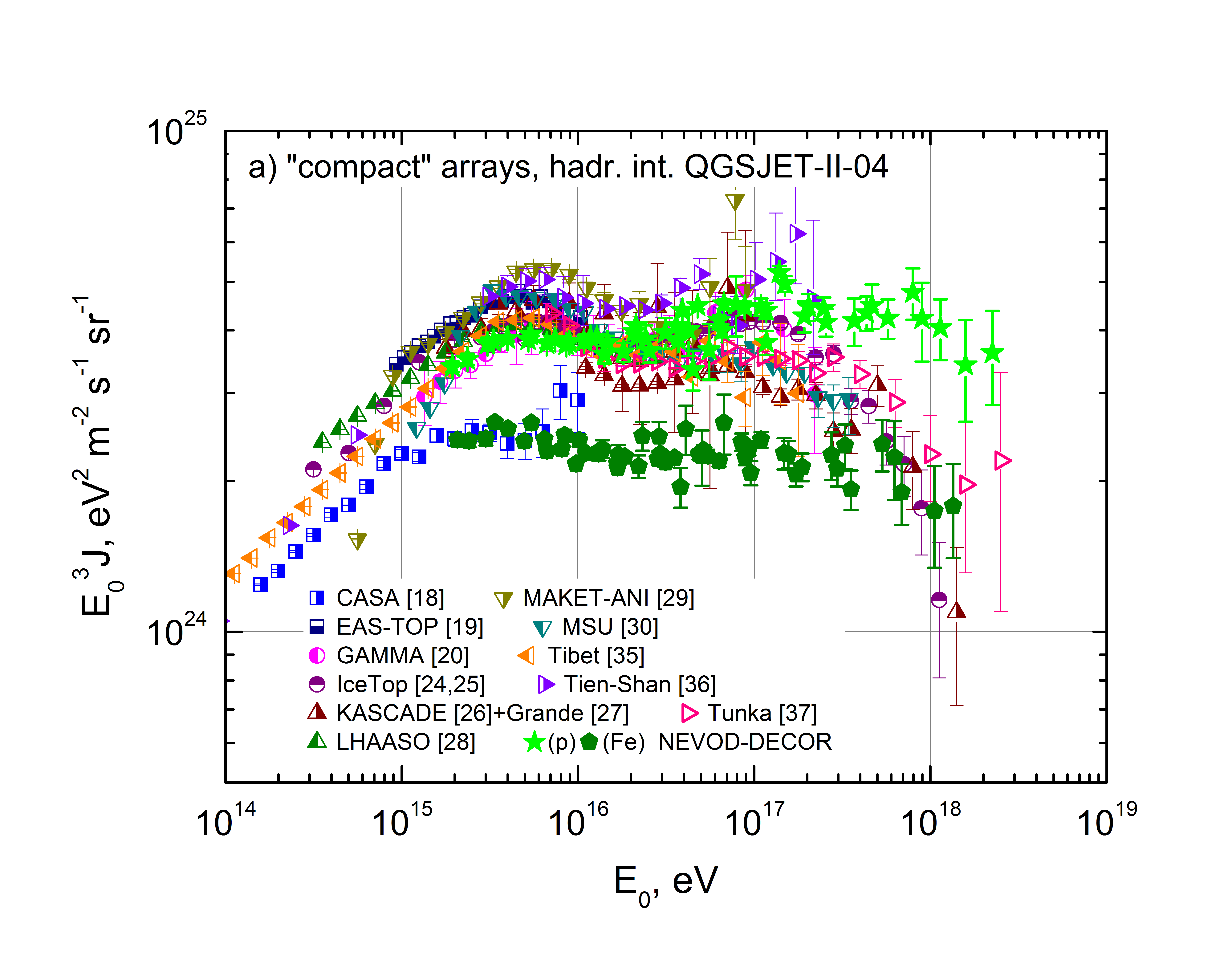}
 \end{minipage}
 \hfill
 \begin{minipage}[h]{0.495\textwidth}
  \includegraphics[trim={1.5cm 1.5cm 2cm 2.2cm}, clip, width=.975\linewidth]{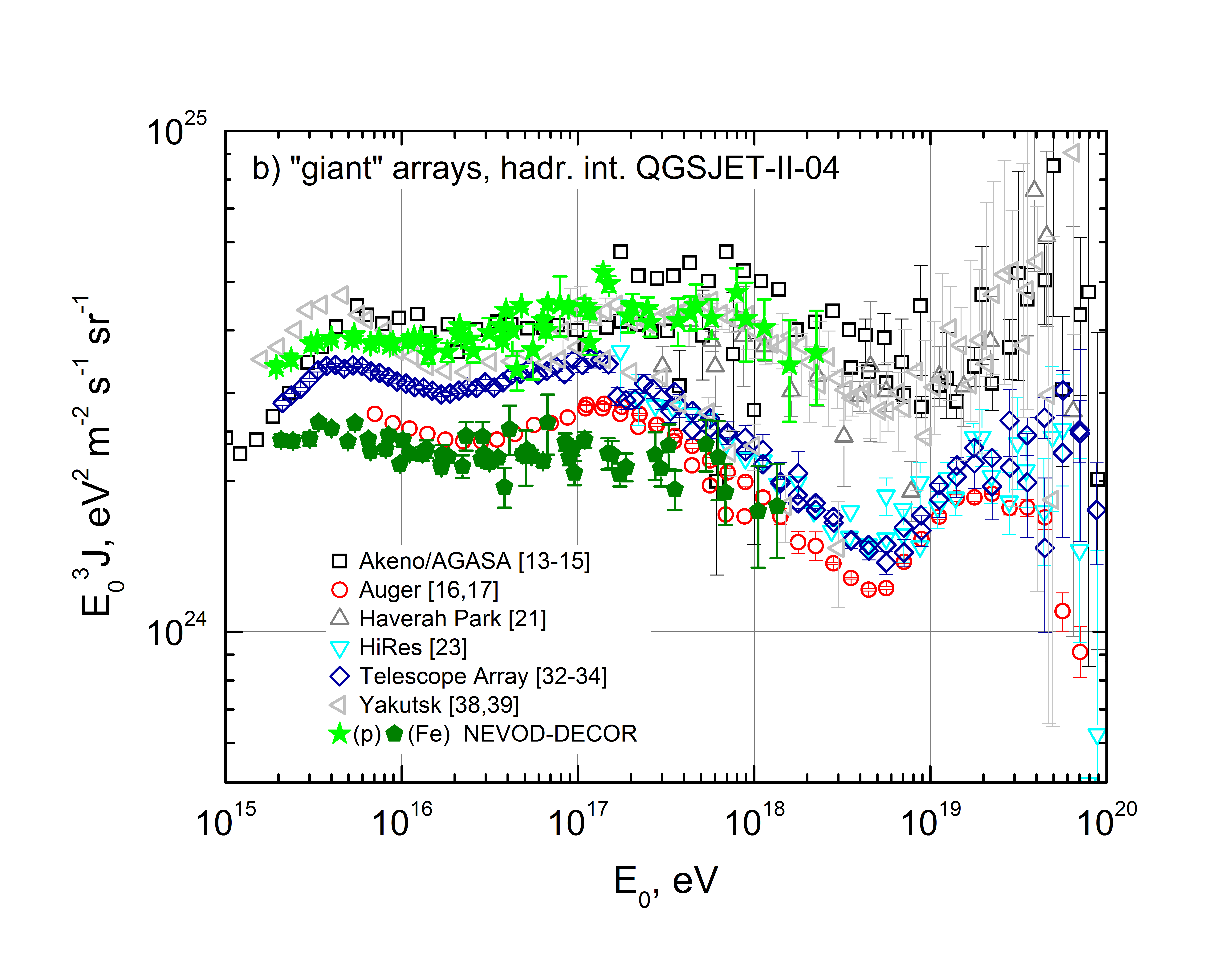}
 \end{minipage}
\caption{\label{fig:14}  The all-particle energy spectrum reconstructed from the NEVOD-DECOR data in comparison with the results of ``compact'' (a) and ``giant'' (b) arrays for studying extensive air showers.}
\end{figure*}

The primary CR energy spectra obtained from the intensity of muon bundles detected by the NEVOD-DECOR, assuming that all primary particles are protons (marked with asterisks) or iron nuclei (pentagons) for the QGSJET-II-04 model of high-energy hadronic interactions, are shown in Fig.~\ref{fig:13}. The values with statistical error for some models are given in Table~\ref{tab:3} of the Appendix. The absolute intensity of the CR flux for showers initiated by iron nuclei is approximately 2 times less than for EAS from protons, since for the same energy and zenith angle the muon density for the former will be higher. The same figure shows the results of a set of experiments~\cite{Akeno1984spe, Akeno1992spe, AGASA2003spe, Auger2021spe, Auger2021speICRC, CASA1999spe, EASTOP1999spe, GAMMA2014spe, HavPark1991spe, HAWC2025spe, HiRes2008spe, IceTop2019spe, IceTop2020spe, KASCADE2005spe, KASCADE2012spe, LHAASO2024spe, MAKETANI2007spe, MSU1991spe, NUCLEON2019spe, TA2018speTALE, TA2023speFD, TA2023speSD, Tibet2008spe, TSh2020spe, Tunka2020spe, Yakutsk2018spe, Yakutsk2019speChe}, in which the reconstruction of the CR energy spectra was carried out based on the measurements of the EAS electron-photon component. In addition, the results of ``compact'' (with an area $\leq 1$ km$^2$) and ''giant'' arrays (with an area $> 1$ km$^2$) are shown on a larger scale in Fig.~\ref{fig:14}a and Fig.~\ref{fig:14}b, respectively.

The intensity of muon bundles (NEVOD-DECOR) at the primary CR energies of $\sim 10^{15}$ – $10^{16}$ eV is in good agreement with the results of the vast majority of experiments, when assuming that the primary cosmic ray particles are protons. The exceptions are the TALE (Telescope Array) and the HEAT (Pierre Auger Observatory) telescopes, which measure mostly Cherenkov radiation from extensive air showers in the atmosphere, as well as the CASA setup, which gives lower estimates of the CR flux intensity than the other experiments. Such estimates in the specified energy range can be consistent with the absolute values of the muon bundle intensity only in the case of a heavy CR mass composition.

At the energies of $\sim 10^{18}$ eV the intensity of muon bundles (NEVOD-DECOR) agrees with the primary CR flux intensity based on measurements of the EAS electron-photon component in the Akeno/AGASA, Haverah Park, and Yakutsk arrays, when assuming a light CR mass composition (protons, He nuclei), or with the data of the High Resolution Fly's Eye, Pierre Auger Observatory, and Telescope Array setups, but under the assumption of a heavy composition (Fe group nuclei). At the same time, the measurements of the depth of shower maximum $X_\text{max}$ in the HiRes, Auger and TA experiments~\cite{HiRes2010xmax,Auger2014xmax,TA2018xmax} indicate a light mass composition of CRs (protons, helium nuclei) at the energies of $\sim 10^{18}$ eV. Let us emphasize that the discrepancy (approximately by a factor of 2) in the CR flux intensity between these two groups of experiments can be related to the approach to the estimation of the primary CR energy. Thus, the first group (Akeno/AGASA, Haverah Park and Yakutsk) uses the density of charged particles in the surface detectors at large distances from the EAS axis (like $\rho_\text{600}$), while the second one (HiRes, Auger, TA) uses the fluorescence technique.

\begin{figure}[hb]
\includegraphics[trim={1.5cm 1.5cm 2cm 2.5cm}, clip, width=.925\linewidth]{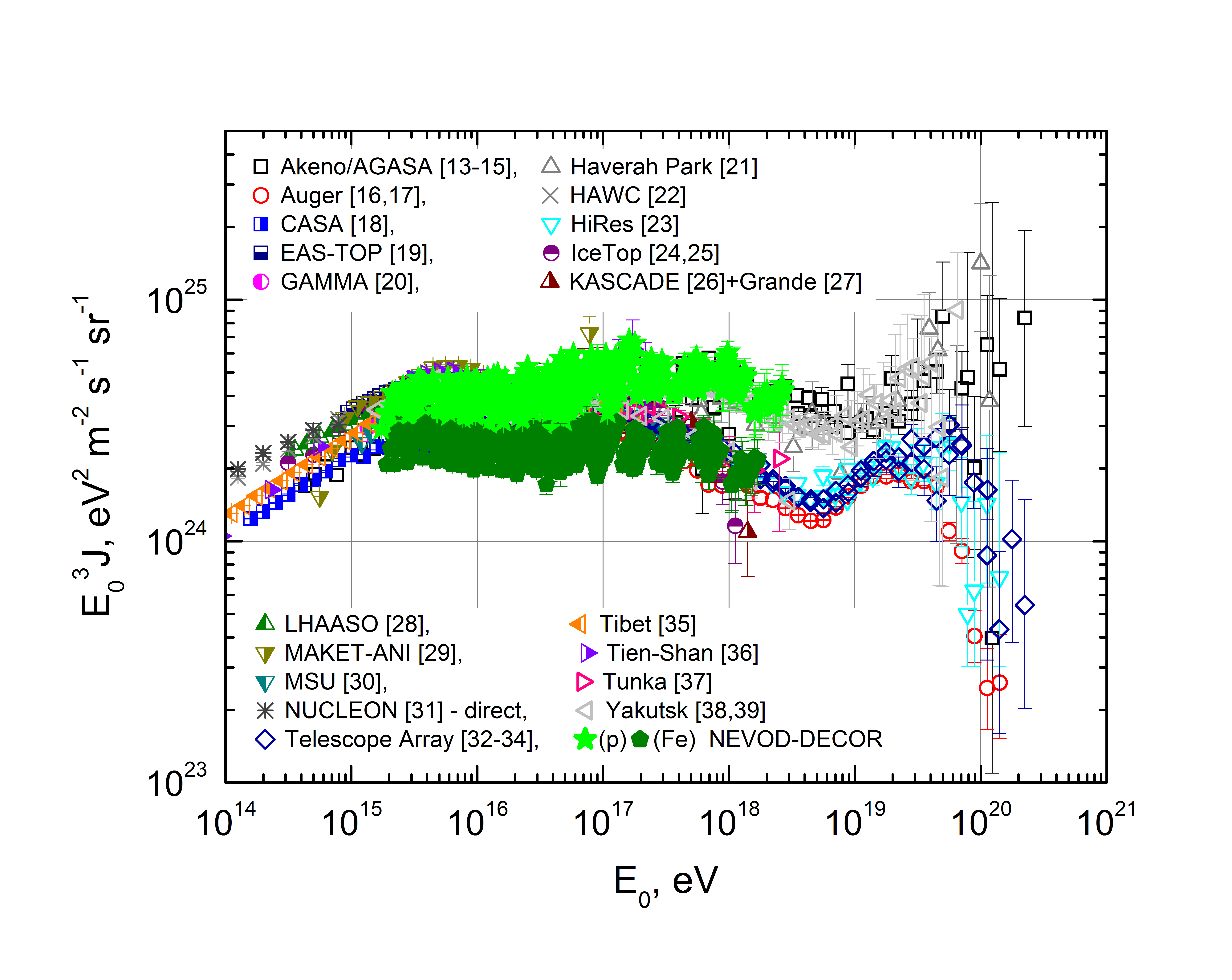}
\caption{\label{fig:15}  The CR energy spectrum according to the NEVOD-DECOR data, assuming that all primary particles are protons (stars) or iron nuclei (pentagons), for six hadronic interaction models (EPOS-LHC, QGSJET-II-04, SIBYLL-2.3c, QGSJET01D, QGSJET-III-01, SIBYLL-2.3e).}
\end{figure}

The uncertainty in the CR flux intensity estimates from the NEVOD-DECOR data, related to the choice of the high-energy hadronic interaction model, including three post-LHC (EPOS-LHC, QGSJET-II-04, SIBYLL-2.3c) models, one pre-LHC (QGSJET01D) model and two new models (QGSJET-III-01, SIBYLL-2.3e) is illustrated in Fig.~\ref{fig:15}. It is about 20$\%$ for primary CR energies $E_0 \sim 10^{15}$ eV, 15$\%$ for $E_0 ~ 10^{18}$ eV and 35$\%$ for $E_0 \sim 10^{16}$ – $10^{17}$ eV when considering EPOS-LHC, QGSJET-II-04, SIBYLL-2.3c, SIBYLL-2.3e models. Taking into account the pre-LHC QGSJET01D model and the new QGSJET-III-01 model, the spread of intensity estimates increases to 25$\%$ for $E_0 ~ 10^{18}$ eV and up to 50$\%$ for $E_0 \sim 10^{15}$ – $10^{17}$ eV. Thus, the primary CR flux intensity estimate from the intensity of muon bundles is more sensitive to extreme assumptions about the mass composition than to the choice of the hadronic interaction model at ultra-high energies.

\section{Behavior of the prinary cosmic ray mass composition}

It is possible to estimate the behavior of the mass composition with a change of the primary CR energy using the NEVOD-DECOR data in the case of an a priori choice of a certain hadronic interaction model and energy spectrum model.

\begin{figure*}[ht]
\includegraphics[trim={1.5cm 1.5cm 1.5cm 1.5cm}, clip, width=.925\linewidth]{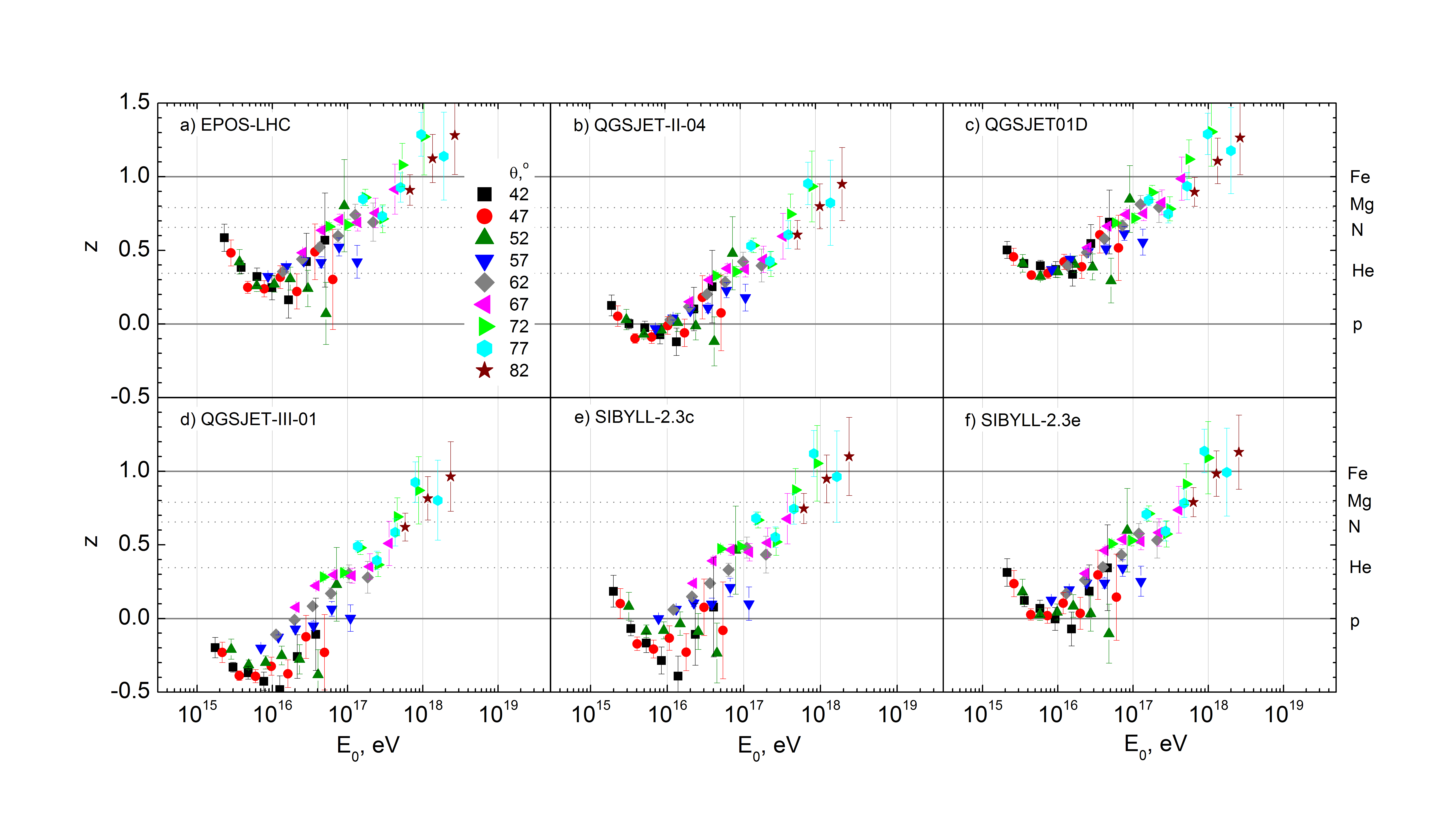}
\caption{\label{fig:16}  The dependencies of \textit{z}-parameter on the energy of primary CRs (protons) derived from the NEVOD-DECOR data (LMDS) for six models of high-energy hadronic interactions and the same model of the CR energy spectrum N$\&$D~\cite{Bogdanov2018}, as defined in Eq. (\ref{eq:7}). The data for nine intervals of the zenith angle are presented with different symbols.}
\end{figure*}

A convenient value that allows obtaining such estimates is the \textit{z}-scale (parameter), which was proposed by the Working group on Hadronic Interactions and Shower Physics (WHISP)~\cite{Dembinski2019whisp} to compare data from various experiments, including NEVOD-DECOR, studying the EAS muon component:
\begin{equation}
z = (\text{ln}N_\mu - \text{ln}N_\mu^\text{p sim})/(\text{ln}N_\mu^\text{Fe sim} - \text{ln}N_\mu^\text{p sim}),
\label{eq:9}
\end{equation}
where $N_\mu$ is the observed value (muon density, muon number, in our case LMDS $dF/dD$, etc.), $N_\mu^\text{p sim}$ and $N_\mu^\text{Fe sim}$ sim are the calculated estimates of this value for EAS formed by primary protons and iron nuclei. Obviously, $z = 0$ or $z = 1$ will mean that the primary cosmic rays consist only of protons or iron nuclei, respectively.

The dependencies of \textit{z}-parameter on the energy of primary CRs (in this case, protons) according to the NEVOD-DECOR data for six models of high-energy hadronic interactions (EPOS-LHC, QGSJET-II-04, QGSJET01D, QGSJET-III-01, SIBYLL-2.3c, SIBYLL-2.3e) and the same model of CR energy spectrum N$\&$D (\ref{eq:7}) are presented in Fig.~\ref{fig:16} and Table~\ref{tab:3} of the Appendix (some of them). The experimental and theoretical LMDS in all nine zenith angle intervals were used to calculate the \textit{z}-parameter. The mean logarithmic values were used as estimates of the effective energies of primary particles (protons) $E_0$ contributing to events with muon bundles.

Note that the \textit{z}-values obtained in different zenith angle ranges from 40$^{\circ}$ to 85$^{\circ}$ overlap and agree well with each other within the errors. Analyzing these dependencies, one can see that the choice of the high-energy hadronic interaction model has a significant effect on the behavior of the primary CR mass composition. For example, for the EPOS-LHC and QGSJET01D models in comparison with the QGSJET-II-04 and SIBYLL-2.3c models, the \textit{z}-parameter is shifted toward a heavier mass composition in the entire PCR energy range: from protons to He nuclei at energies $E_0$ of $\sim 10^{16}$ eV and from Fe nuclei to even heavier nuclei at energies $E_0$ of $\sim 10^{18}$ eV. Besides, for the QGSJET-III-01 model, the \textit{z}-parameter values are negative at energies $E_0$ below $\approx 3 \times 10^{16}$ eV, which corresponds to moderate zenith angles ($\theta =$ 40 – 55$^{\circ}$) of muon bundles.

\begin{figure*}[ht!]
\includegraphics[trim={1.5cm 1.5cm 2cm 1.5cm}, clip, width=.925\linewidth]{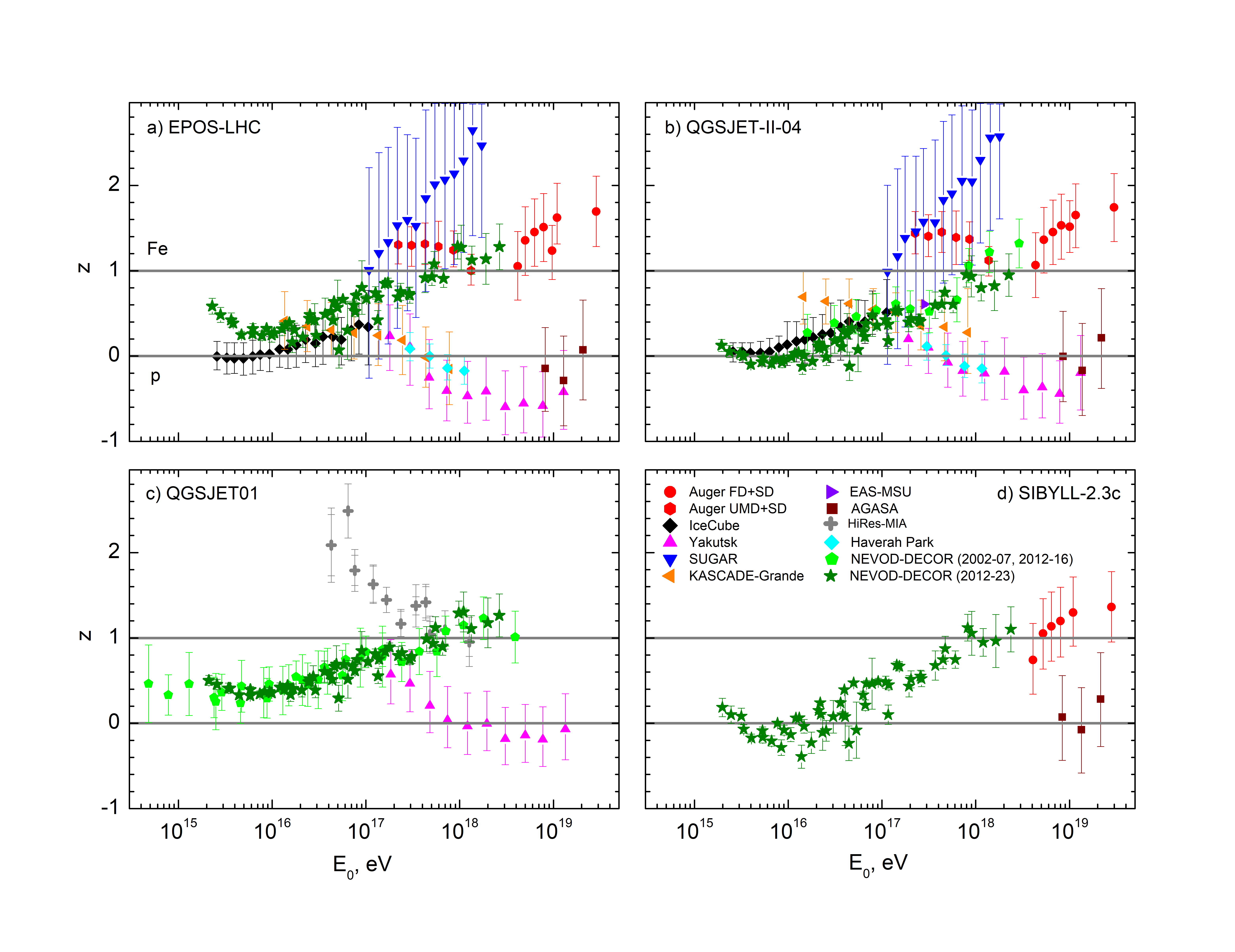}
\caption{\label{fig:17}  The comparison of new NEVOD-DECOR data (2012-2023) on muon bundles converted to the \textit{z}-scale with the early (2002-2007) and combined (2002-2007 and 2012-2016) ones, as well as with the results of measurements of the EAS muon component in other experiments from the WHISP survey~\cite{Arteaga2023whisp}.}
\end{figure*}

A comparison of new NEVOD-DECOR data (2012–2023) on muon bundles (marked with asterisks) with the results of studying the EAS muon component in other experiments (taken from the WHISP analysis~\cite{Arteaga2023whisp}) using the \textit{z}-scale (\ref{eq:9}) is shown in Fig.~\ref{fig:17}. Additionally, Fig.~\ref{fig:17}a and Fig.~\ref{fig:17}c show the early NEVOD-DECOR data (2002 - 2007)~\cite{Bogdanov2010} (pentagons) and the combined 2002 - 2007 and 2012 - 2016 data~\cite{Bogdanov2018} (also pentagons). They are in good agreement with the new ones within the errors, despite some differences in the zenith angle ranges and the models of the PCR energy spectrum, as well as in the selection criteria for events with muon bundles, which confirms the stability of the results obtained with the NEVOD-DECOR setup.

On the whole, from Fig.~\ref{fig:17}, as well as from the WHISP review~\cite{Arteaga2023whisp} where a more complete analysis is presented, it follows that the experiments studying the EAS muon component can be divided into two parts. The data from one part of the setups (HiRes-MIA, Auger, SUGAR, TA) indicate an excess of muons at ultra-high energies compared to model predictions, while the other part of the setups (AGASA, KASCADE-Grande, Haverah Park, Yakutsk) do not see such an excess.

A good agreement between the NEVOD-DECOR data and the IceCube and Auger FD+SD results for the QGSJET-II-04 hadronic interaction model is observed (Fig.~\ref{fig:17}b), despite the fact that their areas differ by orders of magnitude: 70 m$^2$, 1 km$^2$, and 3000 km$^2$, respectively. The \textit{z}-values derived from the NEVOD-DECOR experiment demonstrate a stable growth trend at the primary CR energies above $10^{17}$ eV, and at the energies $\sim 10^{18}$ eV they indicate an extremely heavy (iron group nuclei) CR mass composition.

However, as mentioned above, the expected LMDS (see curves in Fig.~\ref{fig:10}), which are used to calculate the \textit{z}-values based on the NEVOD-DECOR data, depend not only on the shape of the muon LDFs, but also on the choice of the CR energy spectrum model.

\begin{figure}[hb!]
\includegraphics[trim={1.5cm 1.5cm 2cm 2.5cm}, clip, width=.925\linewidth]{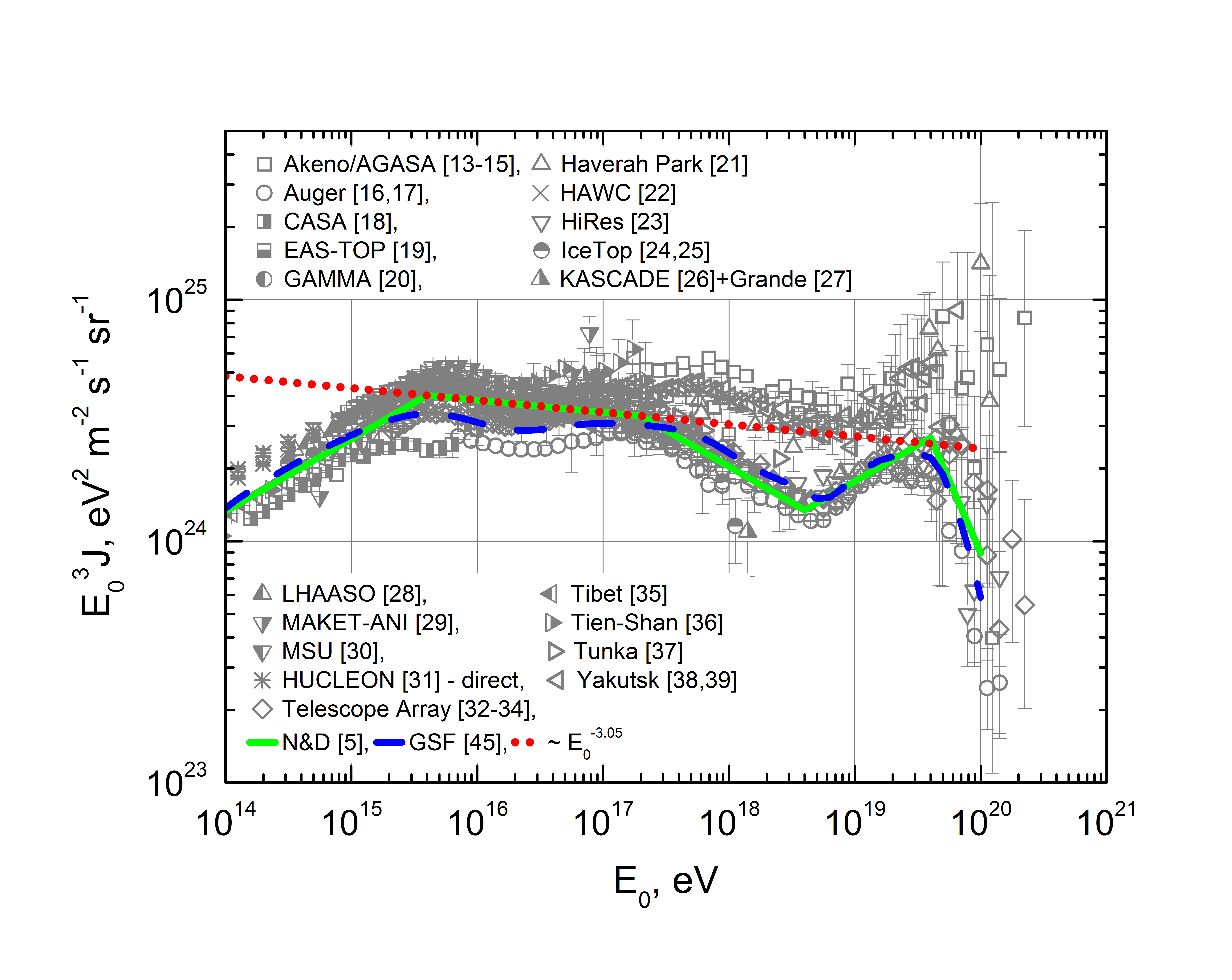}
\caption{\label{fig:18}  The all-particle spectrum as a function of $E_0$ (energy-per-nucleus) from air shower measurements (symbols), and three models of the primary CR energy spectrum (lines).}
\end{figure}

\begin{figure*}[ht!]
\includegraphics[trim={1.5cm 1.5cm 2cm 1.5cm}, clip, width=.9\linewidth]{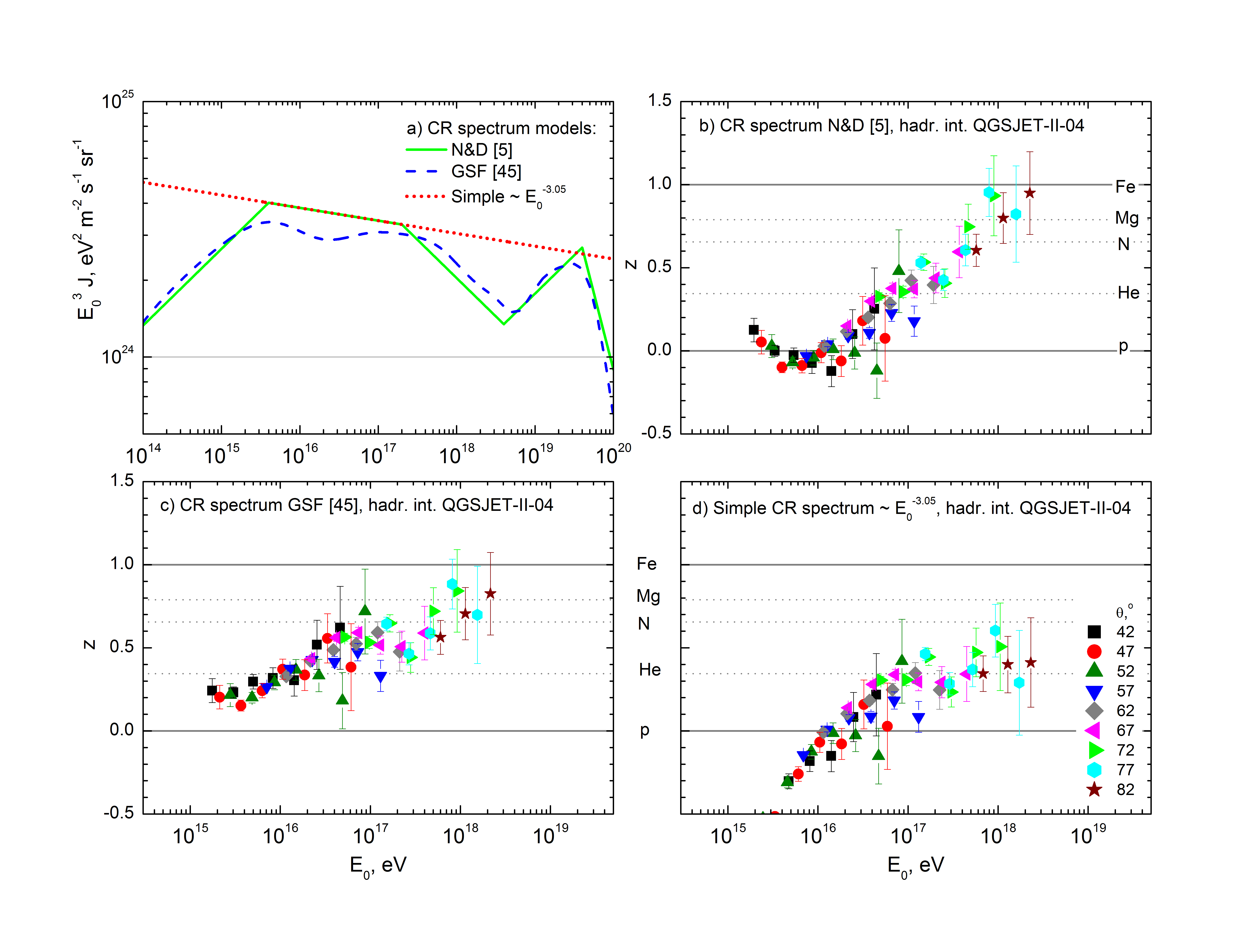}
\caption{\label{fig:19}  Three models of the primary CR energy spectrum (a). The dependencies of the \textit{z}-parameter derived from the NEVOD-DECOR data on energy $E_0$ for these models of the spectrum (b-d) and the same model of high-energy hadronic interactions QGSJET-II-04. The data for 9 intervals of the zenith angle are presented with different symbols.}
\end{figure*}

In order to evaluate the effect of the spectrum model on the \textit{z}-parameter, the Global Spline Fit (GSF)~\cite{Dembinski2017gsf}, used in the WHISP papers, and a simple power-law dependence of the form $dN_2/dE_0 \sim E_0^{-3.05}$ from Eq. (\ref{eq:7}) were chosen. Together with our piecewise power-law approximation N$\&$D (\ref{eq:7}) these models are shown in Fig.~\ref{fig:18} against the background of data of many experiments. They are also plotted separately in Fig.~\ref{fig:19}a by solid, dashed and dotted lines, respectively.

The dependencies of the \textit{z}-parameter on primary energy $E_0$ (Fig.~\ref{fig:19}) for our approximation of the CR spectrum N$\&$D (\ref{eq:7}) and GSF with the same model of high-energy hadronic interactions QGSJET-II-04 differ significantly. In the first case, there is a sharp increase, which corresponds to a change in the CR mass composition from light to heavy (Fig.~\ref{fig:19}b) in the energy range of $10^{16}$ – $10^{18}$ eV. And in the second case, there is a smoother rise from mixed to heavy composition (Fig.~\ref{fig:19}c). This is due to a $10 – 30\%$ discrepancy of intensity in these CR spectrum models. However, in the energy region of $\sim 10^{18}$ eV both spectrum models are based on the EAS fluorescence radiation measurements (HiRes, Auger, TA) and, therefore, both \textit{z}-parameter dependencies on $E_0$ indicate a heavy CR mass composition.
The dependence of the \textit{z}-parameter on $E_0$ for the simple power-law model of the CR energy spectrum $\sim E_0^{-3.05}$ and for the QGSJET-II-04 hadronic interaction model is shown in Fig.~\ref{fig:19}d. Negative values of the \textit{z}-parameter contradict reasonable assumptions about the primary CR mass composition and, therefore, allow us to reject such model of the spectrum in the energy range of $10^{15}$ – $10^{16}$ eV. At the same time at primary particle energies of $\sim 10^{17}$ – $10^{18}$ eV the above spectrum model leads to a mass composition that can be characterized as mixed and slightly varying with energy. Interestingly, the absolute CR flux intensity in this spectrum model lies quite close to the data of the Akeno/AGASA, Haverah Park, and Yakutsk experiments.

\section{Discussion and conclusions}

The long-term experiment on the systematic study of muon bundles in extensive air showers, generated by primary cosmic rays with energies of $10^{15}$ – $10^{19}$ eV, in a wide range of zenith angles was conducted at the NEVOD-DECOR setup. The original approach, the LMDS method, is used to analyze the data. 

The measured characteristics of extensive air showers (and muon bundles) are determined a priori by three unknowns: the energy spectrum of primary cosmic rays, the mass composition, and the hadronic interaction properties. So, the experimental data on the muon bundles intensity allow us: to obtain estimates of the primary CR energy spectrum, making certain assumptions about the mass composition and the model of hadronic interactions; to study the behavior of the CR mass composition using the models of the energy spectrum and hadronic interactions; to test existing hadronic interaction models based on the available data on the energy spectrum and mass composition. It is important to emphasize that the average energies of muons in bundles are hundreds of GeV. Such muons are probably related to the most energetic EAS hadrons, which are formed in the forward kinematic region of interactions.

A comparison of the all-particle energy spectra obtained from two different EAS characteristics showed the following. At primary energies of $\sim 10^{15}$ – $10^{16}$ eV, the NEVOD-DECOR data on muon bundles are well consistent in absolute intensity with the results of most experiments on the electron-photon component under the assumption of a light mass composition. At primary energies of $\sim 10^{18}$ eV, the NEVOD-DECOR data are in good agreement with the estimates of the primary CR flux intensity of the Akeno/AGASA, Haverah Park and Yakutsk arrays under the assumption of a light mass composition (mainly protons), or with the results of the High Resolution Fly's Eye, Pierre Auger Observatory, Telescope Array experiments, but assuming an extremely heavy mass composition (iron nuclei). Perhaps this contradiction is due to the different approaches to estimating the energy of primary particles: in the first case, a measurement of the lateral distribution of shower particles by means of surface detectors is used (or rather, density $\rho_{600}$ at large distances from the shower axis), while in the second case, a measurement of the longitudinal profile of the cascade in the atmosphere by means of fluorescent detectors is used. It should be noted, that the estimates of the CR spectrum based on the NEVOD-DECOR data at energies near $10^{18}$ eV are more sensitive to extreme assumptions about the mass composition (the difference in the flux intensities for protons and iron nuclei is approximately 2 times) than to the choice of the hadronic interaction model (the spread of intensities is about 25$\%$).

Evaluation of the primary CR mass composition according to the NEVOD-DECOR data and their comparison with the results of other experiments exploring the EAS muon component is based on the application of the \textit{z}-parameter proposed by WHISP. If we use for calculations a CR spectrum similar to that obtained by the Akeno/AGASA, Haverah Park, and Yakutsk arrays, our estimates of the mass composition do not show any significant abundance of heavy nuclei at the primary energies of $\sim 10^{18}$ eV. When using the CR spectrum close to that obtained by the HiRes, Pierre Auger Observatory, and Telescope Array experiments, our estimates of the mass composition in this energy range favor extremely heavy (``muon puzzle''), as in a number of other experiments. The overall picture of the mass composition behavior remains the same regardless of the use of a particular interaction model. However, in the latter case, there is a contradiction with the measurements of the EAS maximum depth $X_\text{max}$ in the above-mentioned experiments, and, therefore, a change in the balance between the electron-photon and muon components in the hadronic interaction models is required.

The state-of-art trend in the development of post-LHC models is associated with a shift of the $X_\text{max}$ and with an increase in the muon yield. However, a comparison of the expected LMDS using new models and the NEVOD-DECOR data shows that, although the situation with the description of the energy region of $\sim 10^{18}$ eV corresponding to muon bundles at large zenith angles is improving, the serious discrepancies appear in the energy region of $10^{15}$ – $10^{16}$ eV, corresponding to muon bundles at moderate zenith angles. The most significant difference is observed for the recently released EPOS LHC-R model (with default options) and somewhat less for the QGSJET-III-01 model. To explain the LMDS measured at the NEVOD-DECOR setup within the first of the mentioned models under the assumption of a light CR mass composition at such energies, it is necessary to adjust (decrease by about 1.5 times) the intensity of the primary CR flux obtained from the data of most experiments detecting the EAS electron-photon component, which are consistent with direct measurements on the satellites and balloons.

The cosmic ray research community has some hopes for accelerator experiments to advance understanding of the hadronic interactions and the development of extensive air showers as a consequence. For example, at the CERN's Large Hadron Collider, the collisions of protons with oxygen ions and oxygen with oxygen are carried out, which will make it possible to obtain new data on the generation of secondary particles in the forward kinematic region, on the formation of quark-gluon plasma, etc.

Cosmic ray experiments also play an important role in the study of hadron interactions and in the search for new physical phenomena. In particular, the ``muon puzzle'' may be related to the inclusion of new processes of muon generation at ultra-high energies. Then the key to its solution will be measuring the energy characteristics of the EAS muon component. Preliminary results of the NEVOD-DECOR experiment on measuring the energy deposit of muon bundles have already shown an increase in the average energy of muons in bundles compared to the theory at the primary energies above $10^{17}$ eV.

The commissioning of a new detector TREK~\cite{Zadeba2022} based on the multi-wire drift chambers (total area is of about 250 m$^2$, two track resolution is $\approx 3$ mm) and the upgrade of the Cherenkov water calorimeter NEVOD will allow us to expand the studied primary CR energy range and to improve the accuracy of measuring the number of muons in bundles, as well as to continue studying the energy characteristics of the muon component of air showers. This will provide an opportunity to improve the existing hadronic interaction models or to open the way to the development of new ones.

\begin{acknowledgments}
The work was performed at the Unique Scientific Facility ``Experimental complex NEVOD'' with the financial support provided by the Russian Ministry of Science and Higher Education, project FSWU-2023-0068 ``Fundamental and Applied Research of Cosmic Rays''. Simulations were carried out using the resources of the MEPhI high-performance computing center.
\end{acknowledgments}

\vspace{10pt}
\vspace{10pt}
\vspace{10pt}
\vspace{10pt}
\vspace{10pt}
\vspace{10pt}
\vspace{10pt}

\bibliographystyle{apsrev4-2}
\bibliography{DECOR_MuSpeMass}



\begin{table*}
\section*{Appendix}

\caption{\label{tab:1} The number of events with muon bundles detected by the NEVOD-DECOR setup. The events are sorted into intervals of zenith angle and local density. The ``live'' time of event analysis for zenith angles $\theta \geq 55^{\circ}$ and multiplicity $m \geq 5$ is 75238 h (full statistics), for $\theta =$ 40 – 55$^{\circ}$ and $m \geq 5$ is 6324 h (part of the experimental data), for $\theta =$ 40 – 55$^{\circ}$ and $m = 4$ is 1043 h (marked with the symbol *).}
\begin{ruledtabular}
\begin{tabular}{cccccc}
\multicolumn{2}{c}{$\theta = 42^{\circ}$}& 
\multicolumn{2}{c}{$\theta = 47^{\circ}$}&
\multicolumn{2}{c}{$\theta = 52^{\circ}$} \\ 
\cline{1-2} \cline{3-4} \cline{5-6} \noalign{\smallskip}
$D$, m$^{-2}$ & 
$N$, events & 
$D$, m$^{-2}$ & 
$N$, events & 
$D$, m$^{-2}$ & 
$N$, events \\ 
\hline \noalign{\smallskip}
0.061& 1581$^*$& 0.055& 1411$^*$& 0.051& 1138$^*$ \\
0.112& 6376& 0.102 & 5499& 0.094& 4353 \\
0.198& 3215& 0.181 & 2728& 0.168& 2112 \\
0.338& 1503& 0.308 & 1319& 0.286& 973 \\
0.580& 610& 0.529 & 518& 0.493& 370 \\
1.013& 231& 0.926 & 201& 0.862& 121 \\
1.809& 79& 1.646& 62& 1.532& 53 \\ 
\hline \noalign{\smallskip}
\multicolumn{2}{c}{$\theta = 57^{\circ}$}& 
\multicolumn{2}{c}{$\theta = 62^{\circ}$}&
\multicolumn{2}{c}{$\theta = 67^{\circ}$} \\ 
\cline{1-2} \cline{3-4} \cline{5-6} \noalign{\smallskip}
$D$, m$^{-2}$ & 
$N$, events & 
$D$, m$^{-2}$ & 
$N$, events & 
$D$, m$^{-2}$ & 
$N$, events \\ 
\hline \noalign{\smallskip}
0.088& 35709& 0.084& 20616& 0.081& 10035 \\
0.157& 17283& 0.149& 9751& 0.144& 4739 \\
0.267& 7836& 0.255& 4414& 0.247& 2097 \\
0.460& 3044& 0.439& 1774& 0.427& 768 \\
0.802& 1107& 0.768& 643& 0.751& 258 \\
1.434& 326& 1.370& 194& 1.342& 84 \\ 
\hline \noalign{\smallskip}
\multicolumn{2}{c}{$\theta = 72^{\circ}$}& 
\multicolumn{2}{c}{$\theta = 77^{\circ}$}&
\multicolumn{2}{c}{$\theta = 82^{\circ}$} \\ 
\cline{1-2} \cline{3-4} \cline{5-6} \noalign{\smallskip}
$D$, m$^{-2}$ & 
$N$, events & 
$D$, m$^{-2}$ & 
$N$, events & 
$D$, m$^{-2}$ & 
$N$, events \\ 
\hline \noalign{\smallskip}
0.078& 3751& 0.076& 961& 0.075& 122 \\
0.139& 1620& 0.136& 356& 0.134& 52 \\
0.237& 749& 0.232& 161& 0.231& 22 \\
0.409& 247& 0.399& 72& &  \\
0.717& 97& 0.702& 19& &  \\
1.290& 31& & & &  \\
\end{tabular}
\end{ruledtabular}
\end{table*}

\begin{table*}
\caption{\label{tab:2} The experimental estimates of the local muon density spectra $D^3dF/dD$ followed by their statistical uncertainties for nine zenith angle intervals.}
\begin{ruledtabular}
\begin{tabular}{cccccc}
\multicolumn{2}{c}{$\theta = 42^{\circ}$}& 
\multicolumn{2}{c}{$\theta = 47^{\circ}$}&
\multicolumn{2}{c}{$\theta = 52^{\circ}$} \\
\cline{1-2} \cline{3-4} \cline{5-6} \noalign{\smallskip}
$D$, m$^{-2}$ & 
$D^3dF/dD$, m$^{-4}$s$^{-1}$sr$^{-1}$ & 
$D$, m$^{-2}$ & 
$D^3dF/dD$, m$^{-4}$s$^{-1}$sr$^{-1}$ & 
$D$, m$^{-2}$ & 
$D^3dF/dD$, m$^{-4}$s$^{-1}$sr$^{-1}$ \\ 
\hline \noalign{\smallskip}
0.06& $(2.19 \pm 0.06) \times 10^{-4}$& 0.06& $(1.49 \pm 0.04) \times 10^{-4}$& 0.05& $(9.40 \pm 0.28) \times 10^{-5}$ \\
0.11& $(2.22 \pm 0.03) \times 10^{-4}$& 0.10& $(1.45 \pm 0.02) \times 10^{-4}$& 0.09& $(9.04 \pm 0.14) \times 10^{-5}$ \\
0.20& $(2.22 \pm 0.04) \times 10^{-4}$& 0.18& $(1.44 \pm 0.03) \times 10^{-4}$& 0.17& $(8.82 \pm 0.19) \times 10^{-5}$ \\
0.34& $(2.13 \pm 0.06) \times 10^{-4}$& 0.31& $(1.42 \pm 0.04) \times 10^{-4}$& 0.29& $(8.43 \pm 0.27) \times 10^{-5}$ \\
0.58& $(1.96 \pm 0.08) \times 10^{-4}$& 0.53& $(1.29 \pm 0.06) \times 10^{-4}$& 0.49& $(7.53 \pm 0.39) \times 10^{-5}$ \\
1.01& $(1.98 \pm 0.13) \times 10^{-4}$& 0.93& $(1.31 \pm 0.09) \times 10^{-4}$& 0.86& $(6.35 \pm 0.58) \times 10^{-5}$ \\
1.81& $(1.92 \pm 0.22) \times 10^{-4}$& 1.65& $(1.11 \pm 0.14) \times 10^{-4}$& 1.53& $(7.82 \pm 1.07) \times 10^{-5}$ \\ 
\hline \noalign{\smallskip}
\multicolumn{2}{c}{$\theta = 57^{\circ}$}& 
\multicolumn{2}{c}{$\theta = 62^{\circ}$}&
\multicolumn{2}{c}{$\theta = 67^{\circ}$} \\ 
\cline{1-2} \cline{3-4} \cline{5-6} \noalign{\smallskip}
$D$, m$^{-2}$ & 
$D^3dF/dD$, m$^{-4}$s$^{-1}$sr$^{-1}$ & 
$D$, m$^{-2}$ & 
$D^3dF/dD$, m$^{-4}$s$^{-1}$sr$^{-1}$ & 
$D$, m$^{-2}$ & 
$D^3dF/dD$, m$^{-4}$s$^{-1}$sr$^{-1}$ \\ 
\hline \noalign{\smallskip}
0.09& $(5.18 \pm 0.03) \times 10^{-5}$& 0.08& $(2.55 \pm 0.02) \times 10^{-5}$& 0.08& $(1.09 \pm 0.01) \times 10^{-5}$ \\
0.16& $(5.03 \pm 0.04) \times 10^{-5}$& 0.15& $(2.44 \pm 0.03) \times 10^{-5}$& 0.14& $(1.06 \pm 0.02) \times 10^{-5}$ \\
0.27& $(4.72 \pm 0.05) \times 10^{-5}$& 0.26& $(2.30 \pm 0.04) \times 10^{-5}$& 0.25& $(9.81 \pm 0.21) \times 10^{-6}$ \\
0.46& $(4.27 \pm 0.08) \times 10^{-5}$& 0.44& $(2.15 \pm 0.05) \times 10^{-5}$& 0.43& $(8.49 \pm 0.31) \times 10^{-6}$ \\
0.80& $(4.03 \pm 0.12) \times 10^{-5}$& 0.77& $(2.05 \pm 0.08) \times 10^{-5}$& 0.75& $(7.53 \pm 0.47) \times 10^{-6}$ \\
1.43& $(3.36 \pm 0.19) \times 10^{-5}$& 1.37& $(1.71 \pm 0.12) \times 10^{-5}$& 1.34& $(6.89 \pm 0.75) \times 10^{-6}$ \\ 
\hline \noalign{\smallskip}
\multicolumn{2}{c}{$\theta = 72^{\circ}$}& 
\multicolumn{2}{c}{$\theta = 77^{\circ}$}&
\multicolumn{2}{c}{$\theta = 82^{\circ}$} \\ 
\cline{1-2} \cline{3-4} \cline{5-6} \noalign{\smallskip}
$D$, m$^{-2}$ & 
$D^3dF/dD$, m$^{-4}$s$^{-1}$sr$^{-1}$ & 
$D$, m$^{-2}$ & 
$D^3dF/dD$, m$^{-4}$s$^{-1}$sr$^{-1}$ & 
$D$, m$^{-2}$ & 
$D^3dF/dD$, m$^{-4}$s$^{-1}$sr$^{-1}$ \\ 
\hline \noalign{\smallskip}
0.08& $(3.73 \pm 0.06) \times 10^{-6}$& 0.08& $(8.89 \pm 0.29) \times 10^{-7}$& 0.08& $(1.00 \pm 0.09) \times 10^{-7}$ \\
0.14& $(3.28 \pm 0.08) \times 10^{-6}$& 0.14& $(6.71 \pm 0.36) \times 10^{-7}$& 0.13& $(9.11 \pm 1.26) \times 10^{-8}$ \\
0.24& $(3.20 \pm 0.12) \times 10^{-6}$& 0.23& $(6.37 \pm 0.50) \times 10^{-7}$& 0.23& $(8.03 \pm 1.71) \times 10^{-8}$ \\
0.41& $(2.44 \pm 0.16) \times 10^{-6}$& 0.40& $(6.71 \pm 0.79) \times 10^{-7}$& &  \\
0.72& $(2.54 \pm 0.26) \times 10^{-6}$& 0.70& $(4.66 \pm 1.07) \times 10^{-7}$& &  \\
1.29& $(2.28 \pm 0.41) \times 10^{-6}$& & & &  \\
\end{tabular}
\end{ruledtabular}
\end{table*}

\begin{table*}
\caption{\label{tab:3} The estimates of the primary CR flux intensity $E_0^3J / 10^{24}$ [eV$^2$m$^{-2}$s$^{-1}$sr$^{-1}$] with statistical uncertainties $\pm\sigma_\text{stat}$ for energies $\text{log}_{10}(E_0/\text{eV})$ based on the NEVOD-DECOR data for two extreme assumptions about the CR mass composition (protons and Fe nuclei) and for three models of hadronic interactions (EPOS-LHC, QGSJET-II-04, SIBYLL-2.3e). The values of the \textit{z}-parameter according to the NEVOD-DECOR data for three models of the CR energy spectrum: N$\&$D~\cite{Bogdanov2018} or Eq.(\ref{eq:7}), GSF~\cite{Dembinski2017gsf}, simple power-law $\sim E_0^{-3.05}$ and for three models of hadronic interactions.}
\begin{ruledtabular}
\begin{tabular}{cccccccccc}
\multicolumn{10}{c}{High-energy hadronic interaction model: EPOS-LHC} \\
\hline \noalign{\smallskip} 
&\multicolumn{2}{c}{p}&
\multicolumn{2}{c}{Fe}&
\multicolumn{1}{c}{$\longleftarrow$ N$\&$D~\cite{Bogdanov2018}}&
\multicolumn{2}{c}{GSF~\cite{Dembinski2017gsf}}&
\multicolumn{2}{c}{Simple ($\sim E_0^{-3.05}$)} \\
\cline{2-3} \cline{4-5}  \cline{6-6}  \cline{7-8}  \cline{9-10} \noalign{\smallskip}
$\theta, ^{\circ}$&
$\text{log}_{10}(E_0)$& $E_0^3J \pm \sigma_\text{stat}$&
$\text{log}_{10}(E_0)$& $E_0^3J \pm \sigma_\text{stat}$&
\textit{z}-value $\pm \sigma_\text{stat}$&
$\text{log}_{10}(E_0)$& \textit{z}-value $\pm \sigma_\text{stat}$&
$\text{log}_{10}(E_0)$& \textit{z}-value $\pm \sigma_\text{stat}$ \\
\hline \noalign{\smallskip}
& 15.36& 3.99 $\pm$ 0.10& 15.39& 3.11 $\pm$ 0.08& 0.58 $\pm$ 0.09& 15.32& 0.81 $\pm$ 0.09& 15.25& -0.63 $\pm$ 0.11 \\
& 15.58& 4.43 $\pm$ 0.06& 15.62& 3.36 $\pm$ 0.04& 0.38 $\pm$ 0.04& 15.55& 0.75 $\pm$ 0.04& 15.51& -0.23 $\pm$ 0.05 \\
& 15.79& 4.33 $\pm$ 0.08& 15.82& 3.19 $\pm$ 0.06& 0.32 $\pm$ 0.06& 15.77& 0.81 $\pm$ 0.06& 15.75& 0.06 $\pm$ 0.06 \\
42& 16.00& 4.14 $\pm$ 0.11& 16.02& 3.03 $\pm$ 0.08& 0.25 $\pm$ 0.08& 15.99& 0.82 $\pm$ 0.08& 15.98& 0.16 $\pm$ 0.09 \\
& 16.21& 3.95 $\pm$ 0.16& 16.23& 2.84 $\pm$ 0.12& 0.16 $\pm$ 0.12& 16.22& 0.74 $\pm$ 0.12& 16.22& 0.14 $\pm$ 0.12 \\
& 16.45& 4.21 $\pm$ 0.28& 16.46& 2.99 $\pm$ 0.20& 0.42 $\pm$ 0.19& 16.48& 0.96 $\pm$ 0.19& 16.47& 0.40 $\pm$ 0.19 \\
& 16.70& 4.33 $\pm$ 0.49& 16.70& 3.04 $\pm$ 0.34& 0.57 $\pm$ 0.32& 16.74& 1.02 $\pm$ 0.32& 16.72& 0.52 $\pm$ 0.32 \\
\noalign{\smallskip}
& 15.45& 4.19 $\pm$ 0.11& 15.47& 3.12 $\pm$ 0.08& 0.48 $\pm$ 0.09& 15.41& 0.73 $\pm$ 0.09& 15.35& -0.36 $\pm$ 0.09 \\
& 15.67& 4.32 $\pm$ 0.06& 15.69& 3.10 $\pm$ 0.04& 0.25 $\pm$ 0.04& 15.64& 0.62 $\pm$ 0.04& 15.61& -0.14 $\pm$ 0.04 \\
& 15.89& 4.22 $\pm$ 0.08& 15.90& 2.99 $\pm$ 0.06& 0.24 $\pm$ 0.06& 15.87& 0.71 $\pm$ 0.06& 15.86& 0.09 $\pm$ 0.06 \\
47& 16.10& 4.24 $\pm$ 0.12& 16.10& 2.98 $\pm$ 0.08& 0.32 $\pm$ 0.08& 16.10& 0.84 $\pm$ 0.08& 16.10& 0.28 $\pm$ 0.08 \\
& 16.33& 4.01 $\pm$ 0.18& 16.32& 2.77 $\pm$ 0.12& 0.22 $\pm$ 0.12& 16.34& 0.73 $\pm$ 0.12& 16.33& 0.20 $\pm$ 0.12 \\
& 16.56& 4.32 $\pm$ 0.30& 16.55& 2.98 $\pm$ 0.21& 0.49 $\pm$ 0.19& 16.60& 0.96 $\pm$ 0.19& 16.58& 0.46 $\pm$ 0.19 \\
& 16.80& 3.91 $\pm$ 0.50& 16.79& 2.70 $\pm$ 0.34& 0.30 $\pm$ 0.34& 16.84& 0.68 $\pm$ 0.34& 16.83& 0.23 $\pm$ 0.34 \\
\noalign{\smallskip}
& 15.56& 4.55 $\pm$ 0.14& 15.56& 3.18 $\pm$ 0.09& 0.42 $\pm$ 0.08& 15.52& 0.69 $\pm$ 0.08& 15.48& -0.10 $\pm$ 0.09 \\
& 15.79& 4.34 $\pm$ 0.07& 15.78& 2.96 $\pm$ 0.04& 0.26 $\pm$ 0.04& 15.77& 0.63 $\pm$ 0.04& 15.75& 0.05 $\pm$ 0.04 \\
& 16.02& 4.26 $\pm$ 0.09& 16.00& 2.88 $\pm$ 0.06& 0.27 $\pm$ 0.05& 16.02& 0.71 $\pm$ 0.05& 16.01& 0.21 $\pm$ 0.05 \\
52& 16.24& 4.24 $\pm$ 0.14& 16.21& 2.82 $\pm$ 0.09& 0.31 $\pm$ 0.08& 16.25& 0.77 $\pm$ 0.08& 16.24& 0.29 $\pm$ 0.08 \\
& 16.47& 4.02 $\pm$ 0.21& 16.43& 2.65 $\pm$ 0.14& 0.24 $\pm$ 0.12& 16.50& 0.68 $\pm$ 0.12& 16.48& 0.22 $\pm$ 0.12 \\
& 16.71& 3.64 $\pm$ 0.33& 16.67& 2.38 $\pm$ 0.22& 0.07 $\pm$ 0.21& 16.75& 0.44 $\pm$ 0.21& 16.73& 0.03 $\pm$ 0.21 \\
& 16.96& 4.88 $\pm$ 0.67& 16.91& 3.17 $\pm$ 0.44& 0.80 $\pm$ 0.31& 17.00& 1.08 $\pm$ 0.32& 16.99& 0.72 $\pm$ 0.32 \\
\noalign{\smallskip}
& 15.94& 4.45 $\pm$ 0.02& 15.91& 2.88 $\pm$ 0.02& 0.32 $\pm$ 0.01& 15.93& 0.70 $\pm$ 0.01& 15.92& 0.23 $\pm$ 0.01 \\
& 16.18& 4.49 $\pm$ 0.03& 16.14& 2.85 $\pm$ 0.02& 0.39 $\pm$ 0.02& 16.19& 0.79 $\pm$ 0.02& 16.18& 0.36 $\pm$ 0.02 \\
57& 16.42& 4.48 $\pm$ 0.05& 16.36& 2.81 $\pm$ 0.03& 0.43 $\pm$ 0.02& 16.44& 0.83 $\pm$ 0.02& 16.42& 0.41 $\pm$ 0.02 \\
& 16.65& 4.35 $\pm$ 0.08& 16.59& 2.71 $\pm$ 0.05& 0.42 $\pm$ 0.04& 16.68& 0.77 $\pm$ 0.04& 16.67& 0.39 $\pm$ 0.04 \\
& 16.89& 4.46 $\pm$ 0.13& 16.82& 2.77 $\pm$ 0.08& 0.52 $\pm$ 0.06& 16.93& 0.80 $\pm$ 0.06& 16.92& 0.46 $\pm$ 0.06 \\
& 17.13& 4.16 $\pm$ 0.23& 17.06& 2.55 $\pm$ 0.14& 0.42 $\pm$ 0.11& 17.17& 0.59 $\pm$ 0.11& 17.18& 0.29 $\pm$ 0.11 \\
\noalign{\smallskip}
& 16.14& 4.52 $\pm$ 0.03& 16.08& 2.74 $\pm$ 0.02& 0.36 $\pm$ 0.01& 16.14& 0.72 $\pm$ 0.01& 16.14& 0.33 $\pm$ 0.01 \\
& 16.40& 4.62 $\pm$ 0.05& 16.31& 2.76 $\pm$ 0.03& 0.44 $\pm$ 0.02& 16.42& 0.80 $\pm$ 0.02& 16.40& 0.43 $\pm$ 0.02 \\
62& 16.63& 4.71 $\pm$ 0.07& 16.55& 2.79 $\pm$ 0.04& 0.52 $\pm$ 0.03& 16.66& 0.85 $\pm$ 0.03& 16.65& 0.50 $\pm$ 0.03 \\
& 16.87& 4.79 $\pm$ 0.11& 16.77& 2.83 $\pm$ 0.07& 0.60 $\pm$ 0.04& 16.91& 0.86 $\pm$ 0.05& 16.90& 0.55 $\pm$ 0.04 \\
& 17.10& 5.05 $\pm$ 0.20& 17.01& 2.97 $\pm$ 0.12& 0.74 $\pm$ 0.07& 17.14& 0.92 $\pm$ 0.08& 17.15& 0.64 $\pm$ 0.07 \\
& 17.34& 4.72 $\pm$ 0.34& 17.25& 2.80 $\pm$ 0.20& 0.69 $\pm$ 0.13& 17.38& 0.77 $\pm$ 0.13& 17.41& 0.50 $\pm$ 0.13 \\
\noalign{\smallskip}
& 16.40& 4.86 $\pm$ 0.05& 16.31& 2.74 $\pm$ 0.03& 0.48 $\pm$ 0.02& 16.42& 0.81 $\pm$ 0.02& 16.41& 0.47 $\pm$ 0.02 \\
& 16.66& 5.19 $\pm$ 0.08& 16.57& 2.90 $\pm$ 0.04& 0.64 $\pm$ 0.02& 16.70& 0.93 $\pm$ 0.02& 16.68& 0.61 $\pm$ 0.02 \\
67& 16.90& 5.31 $\pm$ 0.12& 16.80& 2.93 $\pm$ 0.06& 0.71 $\pm$ 0.04& 16.94& 0.94 $\pm$ 0.04& 16.93& 0.66 $\pm$ 0.04 \\
& 17.13& 5.13 $\pm$ 0.18& 17.03& 2.82 $\pm$ 0.10& 0.69 $\pm$ 0.06& 17.18& 0.84 $\pm$ 0.06& 17.19& 0.59 $\pm$ 0.06 \\
& 17.37& 5.03 $\pm$ 0.31& 17.27& 2.84 $\pm$ 0.18& 0.75 $\pm$ 0.10& 17.41& 0.82 $\pm$ 0.10& 17.44& 0.58 $\pm$ 0.10 \\
& 17.63& 4.73 $\pm$ 0.52& 17.52& 2.69 $\pm$ 0.29& 0.91 $\pm$ 0.17& 17.66& 0.89 $\pm$ 0.18& 17.71& 0.61 $\pm$ 0.18 \\
\noalign{\smallskip}
& 16.75& 5.39 $\pm$ 0.09& 16.64& 2.87 $\pm$ 0.05& 0.66 $\pm$ 0.03& 16.79& 0.92 $\pm$ 0.03& 16.78& 0.64 $\pm$ 0.03 \\
& 17.01& 5.31 $\pm$ 0.13& 16.88& 2.80 $\pm$ 0.07& 0.67 $\pm$ 0.04& 17.05& 0.86 $\pm$ 0.04& 17.04& 0.62 $\pm$ 0.04 \\
72& 17.24& 5.89 $\pm$ 0.22& 17.11& 3.07 $\pm$ 0.11& 0.86 $\pm$ 0.06& 17.28& 0.97 $\pm$ 0.06& 17.29& 0.75 $\pm$ 0.06 \\
& 17.47& 4.77 $\pm$ 0.30& 17.35& 2.63 $\pm$ 0.17& 0.71 $\pm$ 0.09& 17.50& 0.74 $\pm$ 0.10& 17.54& 0.51 $\pm$ 0.10 \\
& 17.73& 5.17 $\pm$ 0.53& 17.60& 2.85 $\pm$ 0.29& 1.08 $\pm$ 0.15& 17.75& 1.04 $\pm$ 0.15& 17.81& 0.76 $\pm$ 0.16 \\
& 18.02& 4.84 $\pm$ 0.87& 17.88& 2.67 $\pm$ 0.48& 1.27 $\pm$ 0.26& 18.02& 1.16 $\pm$ 0.27& 18.08& 0.79 $\pm$ 0.28 \\
\noalign{\smallskip}
& 17.20& 6.37 $\pm$ 0.21& 17.06& 3.02 $\pm$ 0.10& 0.85 $\pm$ 0.04& 17.24& 0.96 $\pm$ 0.04& 17.26& 0.77 $\pm$ 0.04 \\
& 17.46& 5.23 $\pm$ 0.28& 17.32& 2.66 $\pm$ 0.14& 0.73 $\pm$ 0.07& 17.50& 0.77 $\pm$ 0.07& 17.53& 0.57 $\pm$ 0.07 \\
77& 17.70& 5.20 $\pm$ 0.41& 17.56& 2.61 $\pm$ 0.21& 0.93 $\pm$ 0.10& 17.73& 0.90 $\pm$ 0.10& 17.79& 0.66 $\pm$ 0.11 \\
& 17.98& 5.72 $\pm$ 0.67& 17.81& 2.91 $\pm$ 0.34& 1.29 $\pm$ 0.15& 17.98& 1.21 $\pm$ 0.15& 18.04& 0.91 $\pm$ 0.16 \\
& 18.28& 4.03 $\pm$ 0.93& 18.11& 2.11 $\pm$ 0.48& 1.14 $\pm$ 0.30& 18.26& 1.01 $\pm$ 0.30& 18.31& 0.57 $\pm$ 0.32 \\
\noalign{\smallskip}
& 17.83& 5.07 $\pm$ 0.46& 17.63& 2.42 $\pm$ 0.22& 0.91 $\pm$ 0.10& 17.85& 0.86 $\pm$ 0.11& 17.90& 0.62 $\pm$ 0.12 \\
82& 18.13& 4.82 $\pm$ 0.67& 17.93& 2.37 $\pm$ 0.33& 1.12 $\pm$ 0.16& 18.12& 1.02 $\pm$ 0.17& 18.17& 0.68 $\pm$ 0.18 \\
& 18.43& 4.22 $\pm$ 0.90& 18.23& 2.18 $\pm$ 0.46& 1.28 $\pm$ 0.27& 18.40& 1.14 $\pm$ 0.27& 18.43& 0.69 $\pm$ 0.28 \\
\end{tabular}
\end{ruledtabular}
\end{table*}

\begin{table*}
\begin{ruledtabular}
\begin{tabular}{cccccccccc}
\multicolumn{10}{c}{High-energy hadronic interaction model: QGSJET-II-04} \\
\hline \noalign{\smallskip} 
&\multicolumn{2}{c}{p}&
\multicolumn{2}{c}{Fe}&
\multicolumn{1}{c}{$\longleftarrow$ N$\&$D~\cite{Bogdanov2018}}&
\multicolumn{2}{c}{GSF~\cite{Dembinski2017gsf}}&
\multicolumn{2}{c}{Simple ($\sim E_0^{-3.05}$)} \\
\cline{2-3} \cline{4-5}  \cline{6-6}  \cline{7-8}  \cline{9-10} \noalign{\smallskip}
$\theta, ^{\circ}$&
$\text{log}_{10}(E_0)$& $E_0^3J \pm \sigma_\text{stat}$&
$\text{log}_{10}(E_0)$& $E_0^3J \pm \sigma_\text{stat}$&
\textit{z}-value $\pm \sigma_\text{stat}$&
$\text{log}_{10}(E_0)$& \textit{z}-value $\pm \sigma_\text{stat}$&
$\text{log}_{10}(E_0)$& \textit{z}-value $\pm \sigma_\text{stat}$ \\
\hline \noalign{\smallskip}
& 15.29& 3.38 $\pm$ 0.08& 15.32& 2.42 $\pm$ 0.06& 0.13 $\pm$ 0.07& 15.24& 0.24 $\pm$ 0.07& 15.16& -1.00 $\pm$ 0.08 \\
& 15.52& 3.80 $\pm$ 0.05& 15.53& 2.62 $\pm$ 0.03& 0.00 $\pm$ 0.03& 15.48& 0.23 $\pm$ 0.03& 15.43& -0.61 $\pm$ 0.04 \\
& 15.73& 3.92 $\pm$ 0.07& 15.74& 2.61 $\pm$ 0.05& -0.03 $\pm$ 0.04& 15.70& 0.30 $\pm$ 0.04& 15.68& -0.30 $\pm$ 0.04 \\
42& 15.93& 3.75 $\pm$ 0.10& 15.93& 2.47 $\pm$ 0.06& -0.07 $\pm$ 0.06& 15.92& 0.32 $\pm$ 0.06& 15.91& -0.18 $\pm$ 0.06 \\
& 16.15& 3.58 $\pm$ 0.14& 16.14& 2.32 $\pm$ 0.09& -0.12 $\pm$ 0.09& 16.15& 0.30 $\pm$ 0.09& 16.15& -0.15 $\pm$ 0.09 \\
& 16.38& 3.84 $\pm$ 0.25& 16.37& 2.46 $\pm$ 0.16& 0.10 $\pm$ 0.15& 16.41& 0.52 $\pm$ 0.15& 16.40& 0.08 $\pm$ 0.15 \\
& 16.63& 4.01 $\pm$ 0.45& 16.61& 2.54 $\pm$ 0.29& 0.25 $\pm$ 0.25& 16.66& 0.62 $\pm$ 0.25& 16.65& 0.22 $\pm$ 0.25 \\
\noalign{\smallskip}
& 15.37& 3.50 $\pm$ 0.09& 15.38& 2.40 $\pm$ 0.06& 0.05 $\pm$ 0.07& 15.33& 0.20 $\pm$ 0.07& 15.26& -0.80 $\pm$ 0.08 \\
& 15.60& 3.85 $\pm$ 0.05& 15.60& 2.55 $\pm$ 0.03& -0.10 $\pm$ 0.03& 15.56& 0.15 $\pm$ 0.03& 15.52& -0.52 $\pm$ 0.03 \\
& 15.82& 3.76 $\pm$ 0.07& 15.81& 2.42 $\pm$ 0.05& -0.09 $\pm$ 0.04& 15.80& 0.24 $\pm$ 0.04& 15.78& -0.26 $\pm$ 0.04 \\
47& 16.03& 3.80 $\pm$ 0.10& 16.01& 2.41 $\pm$ 0.07& -0.01 $\pm$ 0.06& 16.03& 0.37 $\pm$ 0.06& 16.02& -0.07 $\pm$ 0.06 \\
& 16.26& 3.62 $\pm$ 0.16& 16.23& 2.26 $\pm$ 0.10& -0.06 $\pm$ 0.09& 16.27& 0.34 $\pm$ 0.09& 16.26& -0.08 $\pm$ 0.09 \\
& 16.50& 3.95 $\pm$ 0.28& 16.46& 2.45 $\pm$ 0.17& 0.18 $\pm$ 0.15& 16.53& 0.56 $\pm$ 0.15& 16.51& 0.16 $\pm$ 0.15 \\
& 16.74& 3.65 $\pm$ 0.46& 16.70& 2.24 $\pm$ 0.28& 0.08 $\pm$ 0.26& 16.79& 0.38 $\pm$ 0.26& 16.77& 0.03 $\pm$ 0.26 \\
\noalign{\smallskip}
& 15.49& 3.76 $\pm$ 0.11& 15.48& 2.43 $\pm$ 0.07& 0.03 $\pm$ 0.07& 15.44& 0.21 $\pm$ 0.07& 15.39& -0.53 $\pm$ 0.07 \\
& 15.72& 3.83 $\pm$ 0.06& 15.70& 2.40 $\pm$ 0.04& -0.07 $\pm$ 0.03& 15.69& 0.20 $\pm$ 0.03& 15.66& -0.31 $\pm$ 0.03 \\
& 15.95& 3.78 $\pm$ 0.08& 15.91& 2.32 $\pm$ 0.05& -0.04 $\pm$ 0.04& 15.94& 0.29 $\pm$ 0.04& 15.93& -0.13 $\pm$ 0.04 \\
52& 16.17& 3.78 $\pm$ 0.12& 16.12& 2.27 $\pm$ 0.07& 0.01 $\pm$ 0.06& 16.17& 0.37 $\pm$ 0.06& 16.17& -0.01 $\pm$ 0.06 \\
& 16.41& 3.64 $\pm$ 0.19& 16.35& 2.14 $\pm$ 0.11& -0.01 $\pm$ 0.10& 16.43& 0.33 $\pm$ 0.10& 16.42& -0.03 $\pm$ 0.10 \\
& 16.65& 3.33 $\pm$ 0.30& 16.58& 1.95 $\pm$ 0.18& -0.12 $\pm$ 0.17& 16.69& 0.18 $\pm$ 0.17& 16.67& -0.15 $\pm$ 0.17 \\
& 16.90& 4.51 $\pm$ 0.62& 16.83& 2.62 $\pm$ 0.36& 0.48 $\pm$ 0.25& 16.94& 0.72 $\pm$ 0.26& 16.93& 0.42 $\pm$ 0.25 \\
\noalign{\smallskip}
& 15.87& 3.83 $\pm$ 0.02& 15.82& 2.29 $\pm$ 0.01& -0.03 $\pm$ 0.01& 15.85& 0.27 $\pm$ 0.01& 15.84& -0.14 $\pm$ 0.01 \\
& 16.11& 3.87 $\pm$ 0.03& 16.05& 2.27 $\pm$ 0.02& 0.04 $\pm$ 0.01& 16.11& 0.37 $\pm$ 0.01& 16.10& 0.01 $\pm$ 0.01 \\
57& 16.34& 3.88 $\pm$ 0.04& 16.27& 2.24 $\pm$ 0.03& 0.09 $\pm$ 0.02& 16.35& 0.43 $\pm$ 0.02& 16.34& 0.08 $\pm$ 0.02 \\
& 16.57& 3.82 $\pm$ 0.07& 16.50& 2.18 $\pm$ 0.04& 0.11 $\pm$ 0.03& 16.60& 0.42 $\pm$ 0.03& 16.59& 0.09 $\pm$ 0.03 \\
& 16.82& 3.99 $\pm$ 0.12& 16.74& 2.24 $\pm$ 0.07& 0.23 $\pm$ 0.05& 16.86& 0.47 $\pm$ 0.05& 16.85& 0.18 $\pm$ 0.05 \\
& 17.07& 3.78 $\pm$ 0.21& 16.98& 2.08 $\pm$ 0.12& 0.18 $\pm$ 0.09& 17.11& 0.33 $\pm$ 0.09& 17.12& 0.08 $\pm$ 0.09 \\
\noalign{\smallskip}
& 16.07& 3.87 $\pm$ 0.03& 15.99& 2.16 $\pm$ 0.02& 0.03 $\pm$ 0.01& 16.07& 0.33 $\pm$ 0.01& 16.06& -0.01 $\pm$ 0.01 \\
& 16.32& 3.97 $\pm$ 0.04& 16.22& 2.17 $\pm$ 0.02& 0.12 $\pm$ 0.02& 16.34& 0.42 $\pm$ 0.02& 16.33& 0.10 $\pm$ 0.02 \\
62& 16.56& 4.08 $\pm$ 0.06& 16.45& 2.21 $\pm$ 0.03& 0.20 $\pm$ 0.02& 16.59& 0.49 $\pm$ 0.02& 16.57& 0.18 $\pm$ 0.02 \\
& 16.80& 4.19 $\pm$ 0.10& 16.69& 2.26 $\pm$ 0.05& 0.29 $\pm$ 0.04& 16.84& 0.52 $\pm$ 0.04& 16.83& 0.25 $\pm$ 0.04 \\
& 17.04& 4.45 $\pm$ 0.18& 16.93& 2.39 $\pm$ 0.09& 0.42 $\pm$ 0.06& 17.08& 0.59 $\pm$ 0.06& 17.08& 0.35 $\pm$ 0.06 \\
& 17.28& 4.27 $\pm$ 0.31& 17.17& 2.27 $\pm$ 0.16& 0.40 $\pm$ 0.11& 17.33& 0.48 $\pm$ 0.12& 17.35& 0.25 $\pm$ 0.12 \\
\noalign{\smallskip}
& 16.33& 4.08 $\pm$ 0.04& 16.23& 2.13 $\pm$ 0.02& 0.15 $\pm$ 0.02& 16.35& 0.43 $\pm$ 0.02& 16.34& 0.14 $\pm$ 0.02 \\
& 16.59& 4.38 $\pm$ 0.06& 16.47& 2.27 $\pm$ 0.03& 0.30 $\pm$ 0.02& 16.63& 0.56 $\pm$ 0.02& 16.61& 0.28 $\pm$ 0.02 \\
67& 16.83& 4.51 $\pm$ 0.10& 16.71& 2.31 $\pm$ 0.05& 0.38 $\pm$ 0.03& 16.87& 0.59 $\pm$ 0.03& 16.86& 0.34 $\pm$ 0.03 \\
& 17.07& 4.38 $\pm$ 0.16& 16.95& 2.23 $\pm$ 0.08& 0.37 $\pm$ 0.05& 17.11& 0.52 $\pm$ 0.05& 17.12& 0.30 $\pm$ 0.05 \\
& 17.31& 4.45 $\pm$ 0.28& 17.19& 2.26 $\pm$ 0.14& 0.44 $\pm$ 0.09& 17.35& 0.51 $\pm$ 0.09& 17.38& 0.29 $\pm$ 0.09 \\
& 17.57& 4.18 $\pm$ 0.46& 17.44& 2.26 $\pm$ 0.25& 0.60 $\pm$ 0.15& 17.60& 0.59 $\pm$ 0.16& 17.65& 0.34 $\pm$ 0.17 \\
\noalign{\smallskip}
& 16.68& 4.48 $\pm$ 0.07& 16.55& 2.23 $\pm$ 0.04& 0.33 $\pm$ 0.02& 16.72& 0.56 $\pm$ 0.02& 16.70& 0.31 $\pm$ 0.02 \\
& 16.95& 4.45 $\pm$ 0.11& 16.80& 2.19 $\pm$ 0.05& 0.35 $\pm$ 0.03& 16.99& 0.53 $\pm$ 0.04& 16.98& 0.31 $\pm$ 0.03 \\
72& 17.18& 4.95 $\pm$ 0.18& 17.04& 2.42 $\pm$ 0.09& 0.53 $\pm$ 0.05& 17.22& 0.65 $\pm$ 0.05& 17.23& 0.45 $\pm$ 0.05 \\
& 17.41& 4.15 $\pm$ 0.26& 17.27& 2.13 $\pm$ 0.14& 0.41 $\pm$ 0.09& 17.45& 0.44 $\pm$ 0.09& 17.48& 0.23 $\pm$ 0.09 \\
& 17.67& 4.48 $\pm$ 0.46& 17.52& 2.35 $\pm$ 0.24& 0.75 $\pm$ 0.14& 17.70& 0.72 $\pm$ 0.14& 17.76& 0.47 $\pm$ 0.15 \\
& 17.95& 4.22 $\pm$ 0.76& 17.80& 2.23 $\pm$ 0.40& 0.93 $\pm$ 0.24& 17.96& 0.84 $\pm$ 0.25& 18.02& 0.51 $\pm$ 0.26 \\
\noalign{\smallskip}
& 17.14& 5.21 $\pm$ 0.17& 16.97& 2.33 $\pm$ 0.08& 0.53 $\pm$ 0.04& 17.18& 0.64 $\pm$ 0.04& 17.19& 0.46 $\pm$ 0.04 \\
& 17.39& 4.42 $\pm$ 0.23& 17.24& 2.06 $\pm$ 0.11& 0.43 $\pm$ 0.06& 17.43& 0.46 $\pm$ 0.07& 17.46& 0.28 $\pm$ 0.07 \\
77& 17.64& 4.33 $\pm$ 0.34& 17.47& 2.12 $\pm$ 0.17& 0.61 $\pm$ 0.09& 17.67& 0.59 $\pm$ 0.10& 17.72& 0.37 $\pm$ 0.10 \\
& 17.90& 4.76 $\pm$ 0.56& 17.73& 2.37 $\pm$ 0.28& 0.95 $\pm$ 0.14& 17.91& 0.88 $\pm$ 0.15& 17.97& 0.60 $\pm$ 0.16 \\
& 18.20& 3.41 $\pm$ 0.78& 18.02& 1.74 $\pm$ 0.40& 0.82 $\pm$ 0.29& 18.19& 0.70 $\pm$ 0.29& 18.24& 0.29 $\pm$ 0.32 \\
\noalign{\smallskip}
& 17.76& 4.23 $\pm$ 0.38& 17.55& 1.92 $\pm$ 0.17& 0.60 $\pm$ 0.10& 17.78& 0.56 $\pm$ 0.10& 17.83& 0.34 $\pm$ 0.11 \\
82& 18.06& 4.05 $\pm$ 0.56& 17.84& 1.90 $\pm$ 0.26& 0.80 $\pm$ 0.15& 18.06& 0.71 $\pm$ 0.16& 18.11& 0.40 $\pm$ 0.17 \\
& 18.35& 3.61 $\pm$ 0.77& 18.13& 1.78 $\pm$ 0.38& 0.95 $\pm$ 0.25& 18.33& 0.83 $\pm$ 0.25& 18.36& 0.41 $\pm$ 0.27 \\
\end{tabular}
\end{ruledtabular}
\end{table*}

\begin{table*}
\begin{ruledtabular}
\begin{tabular}{cccccccccc}
\multicolumn{10}{c}{High-energy hadronic interaction model: SIBYLL-2.3e} \\
\hline \noalign{\smallskip} 
&\multicolumn{2}{c}{p}&
\multicolumn{2}{c}{Fe}&
\multicolumn{1}{c}{$\longleftarrow$ N$\&$D~\cite{Bogdanov2018}}&
\multicolumn{2}{c}{GSF~\cite{Dembinski2017gsf}}&
\multicolumn{2}{c}{Simple ($\sim E_0^{-3.05}$)} \\
\cline{2-3} \cline{4-5}  \cline{6-6}  \cline{7-8}  \cline{9-10} \noalign{\smallskip}
$\theta, ^{\circ}$&
$\text{log}_{10}(E_0)$& $E_0^3J \pm \sigma_\text{stat}$&
$\text{log}_{10}(E_0)$& $E_0^3J \pm \sigma_\text{stat}$&
\textit{z}-value $\pm \sigma_\text{stat}$&
$\text{log}_{10}(E_0)$& \textit{z}-value $\pm \sigma_\text{stat}$&
$\text{log}_{10}(E_0)$& \textit{z}-value $\pm \sigma_\text{stat}$ \\
\hline \noalign{\smallskip}
& 15.33& 3.61 $\pm$ 0.09& 15.38& 2.86 $\pm$ 0.07& 0.31 $\pm$ 0.09& 15.28& 0.51 $\pm$ 0.10& 15.20& -1.19 $\pm$ 0.12 \\
& 15.55& 4.03 $\pm$ 0.05& 15.59& 3.08 $\pm$ 0.04& 0.12 $\pm$ 0.04& 15.51& 0.46 $\pm$ 0.04& 15.47& -0.61 $\pm$ 0.05 \\
& 15.76& 4.03 $\pm$ 0.07& 15.79& 2.92 $\pm$ 0.05& 0.07 $\pm$ 0.06& 15.73& 0.51 $\pm$ 0.06& 15.71& -0.24 $\pm$ 0.06 \\
42& 15.96& 3.85 $\pm$ 0.10& 15.98& 2.75 $\pm$ 0.07& 0.00 $\pm$ 0.08& 15.95& 0.51 $\pm$ 0.08& 15.94& -0.12 $\pm$ 0.08 \\
& 16.18& 3.66 $\pm$ 0.15& 16.20& 2.57 $\pm$ 0.10& -0.07 $\pm$ 0.11& 16.19& 0.46 $\pm$ 0.12& 16.18& -0.10 $\pm$ 0.12 \\
& 16.42& 3.92 $\pm$ 0.26& 16.42& 2.71 $\pm$ 0.18& 0.19 $\pm$ 0.18& 16.44& 0.69 $\pm$ 0.18& 16.43& 0.16 $\pm$ 0.18 \\
& 16.66& 4.06 $\pm$ 0.46& 16.66& 2.76 $\pm$ 0.31& 0.34 $\pm$ 0.29& 16.70& 0.77 $\pm$ 0.29& 16.68& 0.30 $\pm$ 0.29 \\
\noalign{\smallskip}
& 15.42& 3.80 $\pm$ 0.10& 15.45& 2.87 $\pm$ 0.08& 0.24 $\pm$ 0.09& 15.37& 0.46 $\pm$ 0.09& 15.31& -0.77 $\pm$ 0.10 \\
& 15.64& 4.04 $\pm$ 0.05& 15.66& 2.86 $\pm$ 0.04& 0.03 $\pm$ 0.04& 15.60& 0.36 $\pm$ 0.04& 15.57& -0.43 $\pm$ 0.04 \\
& 15.86& 3.93 $\pm$ 0.08& 15.87& 2.73 $\pm$ 0.05& 0.02 $\pm$ 0.05& 15.84& 0.45 $\pm$ 0.05& 15.83& -0.16 $\pm$ 0.05 \\
47& 16.07& 3.96 $\pm$ 0.11& 16.07& 2.70 $\pm$ 0.07& 0.10 $\pm$ 0.07& 16.07& 0.58 $\pm$ 0.07& 16.06& 0.05 $\pm$ 0.07 \\
& 16.30& 3.76 $\pm$ 0.17& 16.29& 2.51 $\pm$ 0.11& 0.03 $\pm$ 0.11& 16.31& 0.50 $\pm$ 0.11& 16.30& 0.02 $\pm$ 0.11 \\
& 16.53& 4.09 $\pm$ 0.29& 16.52& 2.69 $\pm$ 0.19& 0.30 $\pm$ 0.17& 16.56& 0.72 $\pm$ 0.17& 16.55& 0.27 $\pm$ 0.17 \\
& 16.78& 3.73 $\pm$ 0.47& 16.76& 2.43 $\pm$ 0.31& 0.14 $\pm$ 0.29& 16.82& 0.48 $\pm$ 0.30& 16.80& 0.09 $\pm$ 0.30 \\
\noalign{\smallskip}
& 15.53& 4.06 $\pm$ 0.12& 15.54& 2.91 $\pm$ 0.09& 0.18 $\pm$ 0.09& 15.49& 0.44 $\pm$ 0.09& 15.44& -0.44 $\pm$ 0.09 \\
& 15.76& 3.98 $\pm$ 0.06& 15.76& 2.73 $\pm$ 0.04& 0.03 $\pm$ 0.04& 15.73& 0.39 $\pm$ 0.04& 15.71& -0.23 $\pm$ 0.04 \\
& 15.99& 3.91 $\pm$ 0.09& 15.97& 2.63 $\pm$ 0.06& 0.04 $\pm$ 0.05& 15.98& 0.48 $\pm$ 0.05& 15.97& -0.04 $\pm$ 0.06 \\
52& 16.20& 3.88 $\pm$ 0.12& 16.18& 2.56 $\pm$ 0.08& 0.08 $\pm$ 0.08& 16.21& 0.54 $\pm$ 0.08& 16.20& 0.06 $\pm$ 0.08 \\
& 16.44& 3.70 $\pm$ 0.19& 16.41& 2.41 $\pm$ 0.13& 0.03 $\pm$ 0.12& 16.46& 0.46 $\pm$ 0.12& 16.45& 0.02 $\pm$ 0.12 \\
& 16.68& 3.39 $\pm$ 0.31& 16.64& 2.16 $\pm$ 0.20& -0.10 $\pm$ 0.20& 16.72& 0.25 $\pm$ 0.20& 16.70& -0.14 $\pm$ 0.20 \\
& 16.92& 4.61 $\pm$ 0.63& 16.88& 2.86 $\pm$ 0.39& 0.60 $\pm$ 0.28& 16.97& 0.86 $\pm$ 0.29& 16.96& 0.53 $\pm$ 0.29 \\
\noalign{\smallskip}
& 15.92& 4.10 $\pm$ 0.02& 15.89& 2.62 $\pm$ 0.01& 0.12 $\pm$ 0.01& 15.90& 0.48 $\pm$ 0.01& 15.89& 0.02 $\pm$ 0.01 \\
& 16.15& 4.14 $\pm$ 0.03& 16.11& 2.58 $\pm$ 0.02& 0.20 $\pm$ 0.02& 16.15& 0.58 $\pm$ 0.02& 16.15& 0.17 $\pm$ 0.02 \\
57& 16.38& 4.14 $\pm$ 0.05& 16.33& 2.52 $\pm$ 0.03& 0.24 $\pm$ 0.02& 16.40& 0.61 $\pm$ 0.02& 16.39& 0.23 $\pm$ 0.02 \\
& 16.62& 4.05 $\pm$ 0.07& 16.55& 2.42 $\pm$ 0.04& 0.24 $\pm$ 0.03& 16.66& 0.57 $\pm$ 0.04& 16.64& 0.21 $\pm$ 0.04 \\
& 16.86& 4.17 $\pm$ 0.13& 16.79& 2.47 $\pm$ 0.07& 0.34 $\pm$ 0.06& 16.90& 0.60 $\pm$ 0.06& 16.89& 0.29 $\pm$ 0.06 \\
& 17.10& 3.87 $\pm$ 0.21& 17.03& 2.28 $\pm$ 0.13& 0.25 $\pm$ 0.10& 17.15& 0.41 $\pm$ 0.11& 17.15& 0.14 $\pm$ 0.11 \\
\noalign{\smallskip}
& 16.11& 4.13 $\pm$ 0.03& 16.05& 2.50 $\pm$ 0.02& 0.17 $\pm$ 0.01& 16.11& 0.53 $\pm$ 0.01& 16.10& 0.14 $\pm$ 0.01 \\
& 16.36& 4.24 $\pm$ 0.04& 16.29& 2.49 $\pm$ 0.03& 0.26 $\pm$ 0.02& 16.38& 0.61 $\pm$ 0.02& 16.37& 0.25 $\pm$ 0.02 \\
62& 16.60& 4.35 $\pm$ 0.07& 16.52& 2.52 $\pm$ 0.04& 0.35 $\pm$ 0.03& 16.63& 0.67 $\pm$ 0.03& 16.61& 0.33 $\pm$ 0.03 \\
& 16.84& 4.45 $\pm$ 0.11& 16.74& 2.55 $\pm$ 0.06& 0.43 $\pm$ 0.04& 16.89& 0.69 $\pm$ 0.04& 16.87& 0.39 $\pm$ 0.04 \\
& 17.08& 4.73 $\pm$ 0.19& 16.98& 2.69 $\pm$ 0.11& 0.58 $\pm$ 0.07& 17.12& 0.75 $\pm$ 0.07& 17.12& 0.49 $\pm$ 0.07 \\
& 17.32& 4.46 $\pm$ 0.32& 17.22& 2.54 $\pm$ 0.18& 0.53 $\pm$ 0.12& 17.36& 0.61 $\pm$ 0.13& 17.39& 0.36 $\pm$ 0.13 \\
\noalign{\smallskip}
& 16.37& 4.40 $\pm$ 0.04& 16.29& 2.48 $\pm$ 0.02& 0.31 $\pm$ 0.02& 16.38& 0.63 $\pm$ 0.02& 16.37& 0.29 $\pm$ 0.02 \\
& 16.63& 4.72 $\pm$ 0.07& 16.54& 2.60 $\pm$ 0.04& 0.46 $\pm$ 0.02& 16.66& 0.75 $\pm$ 0.02& 16.65& 0.44 $\pm$ 0.02 \\
67& 16.87& 4.87 $\pm$ 0.11& 16.77& 2.62 $\pm$ 0.06& 0.54 $\pm$ 0.03& 16.91& 0.76 $\pm$ 0.04& 16.90& 0.49 $\pm$ 0.04 \\
& 17.10& 4.74 $\pm$ 0.17& 17.00& 2.51 $\pm$ 0.09& 0.52 $\pm$ 0.06& 17.15& 0.67 $\pm$ 0.06& 17.15& 0.44 $\pm$ 0.06 \\
& 17.34& 4.74 $\pm$ 0.30& 17.24& 2.52 $\pm$ 0.16& 0.58 $\pm$ 0.09& 17.38& 0.65 $\pm$ 0.10& 17.41& 0.42 $\pm$ 0.10 \\
& 17.61& 4.45 $\pm$ 0.49& 17.48& 2.43 $\pm$ 0.27& 0.74 $\pm$ 0.16& 17.64& 0.72 $\pm$ 0.16& 17.69& 0.46 $\pm$ 0.17 \\
\noalign{\smallskip}
& 16.72& 4.93 $\pm$ 0.08& 16.61& 2.59 $\pm$ 0.04& 0.51 $\pm$ 0.02& 16.76& 0.76 $\pm$ 0.03& 16.74& 0.48 $\pm$ 0.03 \\
& 16.98& 4.90 $\pm$ 0.12& 16.86& 2.53 $\pm$ 0.06& 0.53 $\pm$ 0.04& 17.02& 0.71 $\pm$ 0.04& 17.02& 0.48 $\pm$ 0.04 \\
72& 17.21& 5.47 $\pm$ 0.20& 17.09& 2.77 $\pm$ 0.10& 0.71 $\pm$ 0.05& 17.25& 0.82 $\pm$ 0.05& 17.27& 0.61 $\pm$ 0.05 \\
& 17.45& 4.51 $\pm$ 0.29& 17.32& 2.41 $\pm$ 0.15& 0.57 $\pm$ 0.09& 17.48& 0.60 $\pm$ 0.09& 17.52& 0.38 $\pm$ 0.09 \\
& 17.71& 4.87 $\pm$ 0.49& 17.56& 2.59 $\pm$ 0.26& 0.91 $\pm$ 0.14& 17.74& 0.88 $\pm$ 0.14& 17.80& 0.62 $\pm$ 0.15 \\
& 18.00& 4.53 $\pm$ 0.81& 17.84& 2.43 $\pm$ 0.44& 1.09 $\pm$ 0.25& 18.00& 0.99 $\pm$ 0.25& 18.06& 0.64 $\pm$ 0.27 \\
\noalign{\smallskip}
& 17.18& 5.81 $\pm$ 0.19& 17.03& 2.72 $\pm$ 0.09& 0.71 $\pm$ 0.04& 17.22& 0.82 $\pm$ 0.04& 17.23& 0.63 $\pm$ 0.04 \\
& 17.43& 4.83 $\pm$ 0.26& 17.29& 2.40 $\pm$ 0.13& 0.59 $\pm$ 0.07& 17.47& 0.63 $\pm$ 0.07& 17.50& 0.44 $\pm$ 0.07 \\
77& 17.68& 4.78 $\pm$ 0.38& 17.52& 2.38 $\pm$ 0.19& 0.78 $\pm$ 0.10& 17.71& 0.76 $\pm$ 0.10& 17.76& 0.53 $\pm$ 0.11 \\
& 17.95& 5.24 $\pm$ 0.62& 17.78& 2.65 $\pm$ 0.31& 1.14 $\pm$ 0.15& 17.96& 1.06 $\pm$ 0.15& 18.01& 0.77 $\pm$ 0.16 \\
& 18.25& 3.69 $\pm$ 0.85& 18.08& 1.93 $\pm$ 0.44& 0.99 $\pm$ 0.30& 18.23& 0.86 $\pm$ 0.30& 18.27& 0.44 $\pm$ 0.32 \\
\noalign{\smallskip}
& 17.80& 4.81 $\pm$ 0.44& 17.60& 2.21 $\pm$ 0.20& 0.79 $\pm$ 0.10& 17.82& 0.74 $\pm$ 0.10& 17.88& 0.52 $\pm$ 0.11 \\
82& 18.11& 4.58 $\pm$ 0.64& 17.89& 2.16 $\pm$ 0.30& 0.98 $\pm$ 0.15& 18.10& 0.89 $\pm$ 0.16& 18.15& 0.57 $\pm$ 0.17 \\
& 18.41& 4.00 $\pm$ 0.85& 18.19& 2.00 $\pm$ 0.43& 1.13 $\pm$ 0.25& 18.39& 1.00 $\pm$ 0.25& 18.41& 0.58 $\pm$ 0.27 \\
\end{tabular}
\end{ruledtabular}
\end{table*}

\end{document}